\documentclass{article}
\usepackage{bookstyle}
\usepackage{graphicx}
\usepackage{hyperref}
\usepackage{feynmp}
\usepackage{amssymb}

\def\lesssim{\mathrel{\mathpalette\vereq<}}
\def\gtrsim{\mathrel{\mathpalette\vereq>}}
\makeatletter
\def\vereq#1#2{\lower3pt\vbox{\baselineskip1.5pt \lineskip1.5pt
\ialign{$\m@th#1\hfill##\hfil$\crcr#2\crcr\sim\crcr}}}
\makeatother

\begin{document}
\title{Physics Beyond the Standard Model and~Dark~Matter}
\author{Hitoshi Murayama}
\address{Department of Physics, University of California\\
  Berkeley, CA 94720, USA\\
  {\rm and}\\
  Theoretical Physcis Group, Lawrence Berkeley National Laboratory\\
  Berkeley, CA 94720, USA}

\photo{\includegraphics[width=11cm]{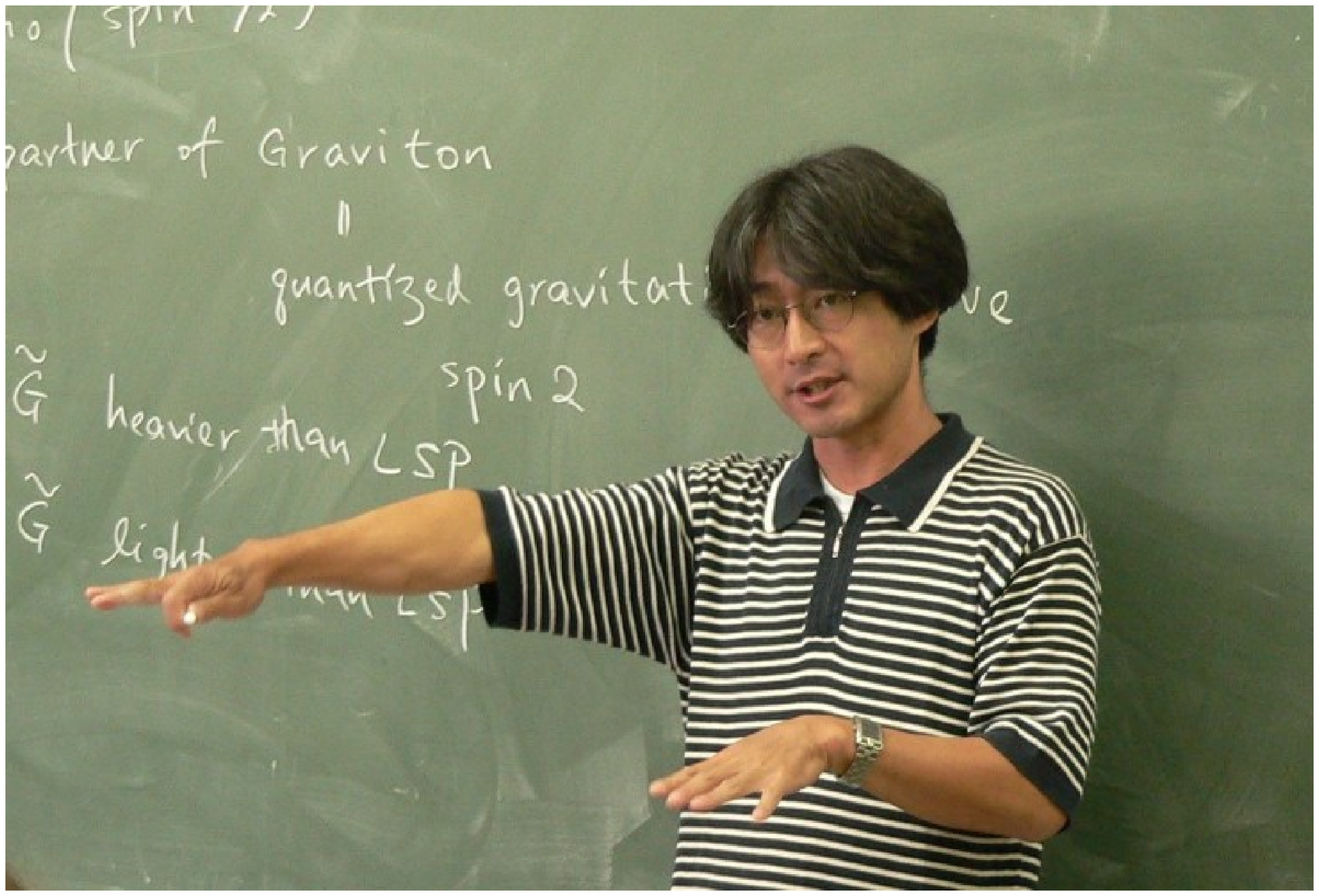}}
\frontmatter
\maketitle    
\mainmatter
%
\section{Introduction}

I'm honored to be invited as a lecturer at a Les Houches summer school
which has a great tradition.  I remember reading many of the lectures
from past summer schools when I was a graduate student and learned a
lot from them.  I'm also looking forward to have a good time in the
midst of beautiful mountains, even though the weather doesn't seem to
be cooperating.  I'm not sure if I will ever get to see Mont Blanc!

I was asked to give four general lectures on physics beyond the
standard model.  This is in some sense an ill-defined assignment,
because it is a vast subject for which we know pratically nothing
about.  It is vast because there are so many possibilities and
speculations, and a lot of ink and many many pages of paper had been
devoted to explore it.  On the other hand, we know practically nothing
about it by definition, because if we did, it should be a part of the
standard model of particle physics already.  I will therefore focus
more on the motivation why we should consider physics beyond the
standard model and discuss a few candidates, and there is no way I can
present all the examples exhaustively.  In addition, after reviewing
the program, I've realized that there are no dedicated lectures on
dark matter.  Since this is a topic where particle physics and
cosmology (I believe) are likely to come together in the near future,
it is relevant to the theme of the school ``Particle Physics and
Cosmology: the Fabric of Spacetime.''  Therefore I will emphasize this
connection in some detail.

Because I try to be pedagogical in lectures, I will probably discuss
many points which some of you already know very well.  Given the wide
spectrum of background you have, I aim at the common denominator.
Hopefully I don't end up boring you all!

\subsection{Particle Physics and Cosmology}

At the first sight, it seems crazy to talk about particle physics and
cosmology together.  Cosmology is the study of the universe, where the
distance scale involved is many Gigaparsecs $\sim 10^{28}$~cm.
Particle physics studies the fundamental constituent of matter, now
reaching the distance scale of $\sim 10^{-17}$~cm.  How can they have
anything in common?

The answer is the Big Bang.  Discovery of Hubble expansion showed that
the visible universe was much smaller in the past, and the study of
cosmic microwave background showed the universe was filled with a hot
plasma made of photons, electrons, and nuclei in thermal equilibrium.
It was {\it hot}\/.  As we contemplate earlier and earlier epochs of
the universe, it was correspondingly smaller and hotter.

On the other hand, the study of small scales $d$ in particle physics
translates to large momentum due to the uncertainty principle, $p \sim
\hbar/d$.  Since large momentum requires relativity, it also means
high energy $E \sim c p \sim \hbar c/d$.  Physics at higher energies
is relevant for the study of higher temperatures $T \sim E/k$, which
was the state of the earlier universe.

This way, Big Bang connects microscopic physics to macroscopic
physics.  And we have already seen two important examples of this
connection.

Atomic and molecular spectroscopy is based on quantum physics at the
atomic distance $d \sim 10^{-8}$~cm.  This spectroscopy is central to
astronomy to identify the chemical composition of faraway stars and
galaxies which we never hope to get to directly and measure their
redshifts to understand their motion including the expansion of the
space itself.  The cosmic microwave background also originates from
the atomic-scale physics when the universe was as hot as $T \sim
4000$~K and hence was in the plasma state.  This is the physics which
we believe we understand from the laboratory experiments and knowledge
of quantum mechanics and hence we expect to be able to extract
interesting information about the universe.  Ironically, cosmic
microwave background also poses a ``wall'' because the universe was
opaque and we cannot ``see'' with photons the state of the universe
before this point.  We have to rely on other kinds of ``messengers''
to extract information about earlier epochs of the universe.

The next example of the micro-macro connection concerns with nuclear
phy\-sics.  The stars are powered by nuclear fusion, obviously a topic
in nuclear physics.  This notion is now well tested by the recent
fantastic development in the study of solar neutrinos, where the core
temperature of the Sun is inferred from the helioseismology and solar
neutrinos which agree at better than a percent level.  Nuclear physics
also determines death of a star.  Relatively heavy stars even end up
with nuclear matter, {\it i.e.}\/ neutron stars, where the entire star
basically becomes a few kilometer-scale nucleus.  On the other hand,
when the universe was as hot as MeV (ten billion degrees Kelvin), it
was too hot for protons and neutrons to be bound in nuclei.  One can
go through theoretical calculations on how the protons and neutrons
became bound in light nuclear species, such as deuterium, $^{3}$He,
$^{4}$He, $^{7}$Li, based on the laboratory measurements of nuclear
fusion cross sections, as well as number of neutrino species from LEP
(Large Electron Positron collider at CERN).  This process is called
Big-Bang Nucleosynthesis (BBN).  There is only one remaining free
parameter in this calculation: cosmic baryon density.  The resulting
predictions can be compared to astronomical determinations of light
element abundances by carefully selecting the sites which are believed
to be not processed by stellar evolutions.  There is (in my humble
opinion) reasonable agreement between the observation and theoretical
predictions (see, {\it e.g.}\/, \cite{Steigman:2005uz}).  This
agreement gives us confidence that we understand the basic history of
the universe since it was as hot as MeV.
\begin{figure}[ht]
  \begin{center}
  \includegraphics[width=0.5\textwidth]{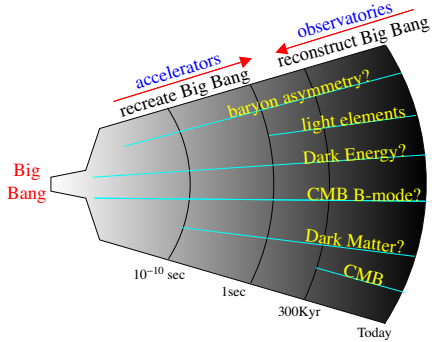}
  \end{center}
  \caption{Possible messengers from early universe.}
  \label{fig:meet}
\end{figure}

We currently do not have messengers from epochs in early universe
above the MeV temperature.  In other words, our understanding of early
universe physics is not tested well for $T \gtrsim $~MeV.  Yet many of
the topics discussed in this school are {\it possible}\/ messengers
from earlier era: dark matter ($10^3$~GeV?), baryon asymmetry of the
universe ($10^{10}$~GeV?), density perturbations (scalar and tensor
components) from the inflationary era ($10^{16}$~GeV?).  These are the
energy scales that laboratory measurements have not reached to reveal
the full particle spectrum and their interactions, hence the realm of
{\it physics beyond the standard model}\/.  Understanding of such
early stages of the universe requires the development in particle
physics, while the universe as a whole may be regarded as a testing
ground of hypothesized particle physics at high energies beyond the
reach of accelerators.  This way, cosmology and particle physics help
and require each other.

\subsection{Next Threshold}

There is a strong anticipation in the community that we are just about
to reveal a new threshold in physics.  Let me tell you why from a
historical perspective.

We (physicists) do not witness crossing a new threshold very often,
but each time it happened, it resulted in a major change in our
understanding of Mother Nature.  

Around year 1900, we crossed the threshold of atomic scale.  It is
impressive to recall how much progress chemists have made without
knowing the underlying dynamics of atoms and molecules.  But the
empirical understanding of chemistry had clear limitation.  For
example, van der Waals equation of state showed there was the distance
scale of about $10^{-8}$~cm below which the state-of-art scientific
knowledge of the time could not be applied, namely the size of atoms.
Once the technology improved to study precision spectroscopy that
allowed people to probe physics inside the atoms, a revolution
followed.  It took about three decades for quantum mechanics to be
fully developed but it forever changed our understanding of nature.
The revolution went on well into the 40's when the marriage of quantum
mechanics and relativity was completed in Quantum ElectroDynamics.

Next important threshold was crossed around 1950 when new hadron
resonances and strange particles were discovered, crossing the
threshold of the strong interaction scale $\sim 10^{-13}$~cm.
Discovery of a zoo of ``elementary particles'' led to a great deal of
confusion for about three decades.  It eventually led to the
revelation of non-perturbative dynamics of quantum field theory,
namely confinement of quarks, dimensional transmutation, and dynamical
symmetry breaking of chiral symmetry.  More importantly, it showed a
new layer in nature where quarks and gluons take over the previous
description of subatomic world with protons and neutrons.  The
experimental verification of this theory, Quantum ChromoDynamics, took
well into 90's at numerous accelerators PETRA, PEP, TRISTAN, LEP, and
HERA.

One more force that is yet to be fully understood is the weak
interaction.  Its scale was known from the time of Fermi back in 1933
when he wrote the first theory of nuclear beta decay.  The theory
contained one dimensionful constant $G_F \approx (300~\mbox{GeV})^{-2}
\approx (10^{-16}~\mbox{cm})^2$.  Seven decades later, we are just
about to reach this energy scale in accelerator experiments, at
Tevatron and LHC.  We do not really know what Nature has in store for
us, but at least we've known all along that this is another important
energy scale in physics.  If we are not misled, this is the energy
scale associated with the cosmic superconductor.  Just like the
Meissner effect lets magnetic field penetrate into a superconductor
only over a finite distance, the cosmic superconductor lets the weak
force carried by $W$ and $Z$ bosons go over a tiny distance: a
billionth of a nanometer.  Right now we are only speculating what
revolution may take place at this distance scale.  A new layer of
matter?  New dimensions of space?  Quantum dimensions?  Maybe string
theory?  We just don't know yet.

Of course historical perspective does not guarantee that history
repeats itself in an equally exciting fashion.  But from all what we
know, there is a good reason to think that indeed a new threshold is
waiting to be discovered at the TeV energy scale, as I will discuss in
the next section.  Another simple fact is that crossing a new
threshold is something like twice-in-a-century experience.  I'm
excited to think that we are just about to witness one, a historic
moment.

An interesting question is what fundamental physics determines these
thresholds.  The atomic scale, that looked like a fundamental
limitation in understanding back in the 19th century, did not turn out
to be a fundamental scale at all.  It is a derived scale from the mass
of the electron and the fundamental constants,
\begin{equation}
  a_B = \frac{\hbar^2}{e^2 m_e} \approx 10^{-8}~\mbox{cm}.
\end{equation}
The strong-interaction scale is also a derived energy scale from the
coupling constant
\begin{equation}
  a_s = M e^{-8\pi^2/g_s^2(M) b_0} \approx 10^{-13}~\mbox{cm},
\end{equation}
where $g_s$ is the strong coupling constant defined at a high-energy
scale $M$ and $b_0$ is the beta function coefficient.  Because of the
asymptotic freedom, the strong coupling constant is weak (what an
oxymoron!) at high energies, while it becomes infinitely strong at low
energies.  The scale of strong interaction is where the strength of
the interaction blows up.  In other words, the two thresholds we have
crossed so far were extremely exciting, yet they turned out to be not
fundamental!  They point to yet deeper physics that determine these
parameters in nature.  Maybe the weak-interaction scale is also a
derived scale from some deeper physics at yet shorter distances.

\section{Why Beyond the Standard Model}

\subsection{Empirical Reasons\label{sec:empirical}}

Until about ten years ago, particle physicists lamented that the
standard model described every new data that came out from experiments
and we didn't have a clue what may lie beyond the standard model.
Much of the discussions on physics beyond the standard model therefore
were not based on data, but rather on theoretical arguments, primarily
philosophical and aesthetic displeasure with the standard model.  It
all changed the last ten years when empirical evidence appeared that
demonstrated that the standard model is incomplete:
\begin{itemize}
\item Non-baryonic dark matter,
\item Dark energy,
\item Neutrino mass,
\item Nearly scale-invariant, Gaussian, and apparently acausal density
  perturbations, 
\item Baryon asymmetry.
\end{itemize}
I will discuss strong evidence for non-baryonic dark matter and dark
matter later in my lectures.  Density fluctuation is covered in many
other lectures in this school by Lev Kofman, Sabino Matarrese, Yannick
Mellier, Simon Prunet, and Romain Teyssier.  Neutrino mass is
discussed by Sergio Pastor, and baryon asymmetry by Jim Cline.  The
bottom line is simple: we already {\it know}\/ that there must be
physics beyond the standard model.  However, we don't necessarily know
the energy (or distance) scale for this new physics, nor what form it
takes.  One conservative approach is to try to accommodate all of
these established empirical facts into the standard model with minimum
particle content: The New Minimal Standard Model
\cite{Davoudiasl:2004be}.  I will discuss some aspects of the model
later.  But theoretical arguments suggest the true model be much
bigger, richer, and more interesting.

\subsection{Philosophical and Aesthetic Reasons\label{sec:philosophical}}

What are the theoretical arguments that demand physics beyond the
standard model?  As I mentioned already, they are based on somewhat
philosophical arguments and aesthetic desires and not exactly on firm
footing.  Nonetheless they are useful and suggestive, especially
because nature did solve some of the similar problems in the past by
invoking interesting mechanisms.  A partial list relevant to my
lectures here is
\begin{itemize}
\item Hierarchy problem: why $G_F \sim 10^{-5}~\mbox{GeV}^{-2} \ll G_N
  \sim 10^{-38}~\mbox{GeV}^{-2}$? 
\item Why $\theta_{QCD} \lesssim 10^{-10} \ll 1$?
\item Why are there three generations of particles?
\item Why are the quantum numbers of particles so strange, yet do
  anomalies cancel so non-trivially?
\end{itemize}
For an expanded list of the ``big questions'', see {\it e.g.}\/,
\cite{Murayama:2003ix}.  

To understand what these questions are about, it is useful to remind
ourselves how the standard model works.  It is a gauge theory based on
the $SU(3) \times SU(2) \times U(1)$ gauge group with the Lagrangian
\begin{eqnarray}
  \lefteqn{
    {\cal L}_{SM} = -\frac{1}{4g^{\prime 2}} B_{\mu\nu} B^{\mu\nu}
    - \frac{1}{2g^2} {\rm Tr} (W_{\mu\nu} W^{\mu\nu})
    - \frac{1}{2g_s^2} {\rm Tr} (G_{\mu\nu} G^{\mu\nu})
  } \nonumber \\
  & & 
  + \bar{Q}_i i{\not\!\! D} Q_i
  + \bar{L}_i i{\not\!\! D} L_i
  + \bar{u}_i i{\not\!\! D} u_i
  + \bar{d}_i i{\not\!\! D} d_i
  + \bar{e}_i i{\not\!\! D} e_i \nonumber \\
  & &
  + (Y_u^{ij} \bar{Q}_i u_j \tilde{H}
  + Y_d^{ij} \bar{Q}_i d_j H
  + Y_l^{ij} \bar{L}_i e_j H + {\rm h.c.}) \nonumber \\
  & & 
  + (D_\mu H)^\dagger (D^\mu H)
  - \lambda (H^\dagger H)^2 - m^2 H^\dagger H
  + \frac{\theta}{32\pi^2} \epsilon^{\mu\nu\rho\sigma} {\rm Tr} (G_{\mu\nu}
  G_{\rho\sigma}). \nonumber \\
  \label{eq:LSM}
\end{eqnarray}
It looks compact enough that it should fit on a T-shirt.\footnote{It
  reminds me of an anecdote from when the standard model was just
  about getting off the ground around 1978.  There was a convergence
  of the data to the standard model and people got very excited about
  it.  Then Tini Veltman gave a talk asking ``do you really think this
  is great model?'' and wrote down every single term in the Lagrangian
  without using a compact notation used here over pages and pages of
  transparencies.  Unfortunately I don't remember who told me this
  story. }  Why don't we see such a T-shirt while we see Maxwell
equations a lot?

The first two lines describe the gauge interactions.  The covariant
derivatives ${\not\!\! D} = \gamma^\mu D_\mu$ in the second line are
determined by the gauge quantum numbers given in this table:

\noindent
\begin{center}
\begin{tabular}{|c|c|c|c|c|}
  \hline
  & $SU(3)$ & $SU(2)$ & $U(1)$ & chirality\\ \hline
  $Q$ & 3 & 2 & $+1/6$ & left\\
  $U$ & 3 & 1 & $+2/3$ & right\\
  $D$ & 3 & 1 & $-1/3$ & right\\
  $L$ & 1 & 2 & $-1/2$ & left\\
  $E$ & 1 & 1 & $-1$ & right\\ \hline
\end{tabular}
\end{center}

\noindent This part of the Lagrangian is well tested, especially by
the LEP/SLC data in the 90's.  However, the quantum number assignments
(especially $U(1)$ hypercharges) appear very strange and actually hard
to remember.\footnote{I often told my friends that I chose physics
  over chemistry or biology because I didn't want to memorize
  anything, but this kind of table casts serious doubt on my choice!}
Why this peculiar assignment is one of the things people don't like
about the standard model.  In addition, they are subject to
non-trivial anomaly cancellation conditions for $SU(3)^2 U(1)$,
$SU(2)^2 U(1)$, ${\rm gravity}^2 U(1)$, $U(1)^3$, and Witten's $SU(2)$
anomalies.  Many of us are left with the feeling that there must be a
deep reason for this baroque quantum number assignments which had led
to the idea of grand unification.

The third line of the Lagrangian comes with the generation index
$i,j=1,2,3$ and is responsible for masses and mixings of quarks and
masses of charged leptons.  The quark part has been tested precisely
in this decade at $B$-factories while there is a glaring omission of
neutrino masses and mixings that became established since 1998.  In
addition, it appears unnecessary for nature to repeat elementary
particles three times.  The repetition of generations and the origin
of mass and mixing patterns remains an unexplained mystery in the
standard model.

The last line is completely untested.  The first two terms describe
the Higgs field and its interaction to the gauge fields and itself.
Having not seen the Higgs boson so far, it is far from established.
The mere presence of the Higgs field poses an aesthetic problem.  It
is the only spinless field in the model, but it is introduced for the
purpose of doing the most important job in the model.  In addition, we
have not seen {\it any}\/ elementary spinless particle in nature!
Moreover, the potential needs to be chosen with $m^2 < 0$ to cause the
cosmic superconductivity which does not give any reason why our
universe is in this state.  I will discuss more problems about it in a
few minutes.  Overall, this part of the model looks very artificial.

The last term is the so-called $\theta$-term in QCD and violates $T$
and $CP$.  The vacuum angle $\theta$ is periodic under $\theta
\rightarrow \theta + 2\pi$, and hence a ``natural'' value of $\theta$
is believed to be order unity.  On the other hand, the most recent
experimental upper limit on the neutron electric dipole moment $|d_n|
< 2.9 \times 10^{-26} e\mbox{ cm}$ (90\% CL)
\cite{Baker:2006ts}\footnote{This is an amazing limit.  If you blow up
  the neutron to the size of the Earth, this limit corresponds to a
  possible displacement of an electron by less than ten microns. }
translates to a stringent upper limit $\theta < (1.2 \pm 0.6) \times
10^{-10}$ using the formula in \cite{Pospelov:2005pr}.  Why $\theta$
is so much smaller than the ``natural'' value is the strong CP
problem, and again the standard model does not offer any explanations.

Now we have more to say about the Higgs sector (the third line).
Clearly it is very important because (1) this is the only part of the
Standard Model which has a dimensionful parameter and hence sets the
overall energy scale for the model, and (2) it has the effect of
causing cosmic superconductivity without explaining its microscopic
mechanism.  For the usual superconductors studied in the laboratory,
we can use the same Lagrangian, but it is {\it derived}\/ from the
more fundamental theory by Bardeen, Cooper, and Schrieffer.  The weak
attractive force between electrons by the phonon exchange causes
electrons to get bound and condense.  The ``Higgs'' field is the
Cooper pair of electrons.  And one can show why it has this particular
potential.  In the standard model, we do not know if Higgs field is
elementary or if it is made of something else, nor what mechanism
causes it to have this potential.

All the puzzles raised here (and more) cry out for a more fundamental
theory underlying the Standard Model.  What history suggests is that
the fundamental theory lies always at shorter distances than the
distance scale of the problem.  For instance, the equation of state of
the ideal gas was found to be a simple consequence of the statistical
mechanics of free molecules.  The van der Waals equation, which
describes the deviation from the ideal one, was the consequence of the
finite size of molecules and their interactions.  Mendeleev's periodic
table of chemical elements was understood in terms of the bound
electronic states, Pauli exclusion principle and spin.  The existence
of varieties of nuclide was due to the composite nature of nuclei made
of protons and neutrons.  The list could go on and on.  Indeed,
seeking answers at more and more fundamental level is the heart of the
physical science, namely the reductionist approach.

The distance scale of the Standard Model is given by the size of the 
Higgs boson condensate $v = 250$~GeV.  In natural units, it gives 
the distance scale of $d = \hbar c/v = 0.8 \times 10^{-16}$~cm.  We 
therefore would like to study physics at distance scales shorter than 
this eventually, and try to answer puzzles whose partial list was 
given in the previous section.

Then the idea must be that we imagine the Standard Model to be valid 
down to a distance scale shorter than $d$, and then new physics will 
appear which will take over the Standard Model.  But applying the 
Standard Model to a distance scale shorter than $d$ poses a serious 
theoretical problem.  In order to make this point clear, we first 
describe a related problem in the classical electromagnetism, and 
then discuss the case of the Standard Model later along the same 
line \cite{INS}.

\subsection{Positron Analogue\label{sec:positron}}

In the classical electromagnetism, the only dynamical degrees of 
freedom are electrons, electric fields, and magnetic fields.  When an 
electron is present in the vacuum, there is a Coulomb electric field 
around it, which has the energy of
\begin{equation}
    \Delta E_{\rm Coulomb} = \frac{1}{4\pi \varepsilon_{0}}\frac{e^{2}}{r_{e}}.
    \label{eq:ECoulomb}
\end{equation}
Here, $r_{e}$ is the ``size'' of the electron introduced to cutoff the
divergent Coulomb self-energy.  Since this Coulomb self-energy is
there for every electron, it has to be considered to be a part of the
electron rest energy.  Therefore, the mass of the electron receives an
additional contribution due to the Coulomb self-energy:
\begin{equation}
    (m_{e} c^{2})_{\it obs} = (m_{e}c^{2})_{\it bare} + \Delta E_{\rm 
    Coulomb}.
    \label{eq:self}
\end{equation}
Experimentally, we know that the ``size'' of the electron is small,
$r_{e} \lesssim 10^{-17}$~cm.  This implies that the self-energy
$\Delta E$ is greater than 10~GeV or so, and hence the ``bare''
electron mass must be negative to obtain the observed mass of the
electron, with a fine cancellation like\footnote{Do you recognize
  $\pi$?}
\begin{equation}
    0.000511 = (-3.141082 + 3.141593)~{\rm GeV}.
\end{equation}
Even setting a conceptual problem with a negative mass electron aside, 
such a fine-cancellation between the ``bare'' mass of the electron and 
the Coulomb self-energy appears ridiculous.  In order for such a 
cancellation to be absent, we conclude that the classical 
electromagnetism cannot be applied to distance scales shorter than 
$e^{2}/(4\pi \varepsilon_{0} m_{e} c^{2}) = 2.8\times 10^{-13}$~cm.  
This is a long distance in the present-day particle physics' standard.

\begin{figure}[h]
        \centerline{\includegraphics[scale=0.5]{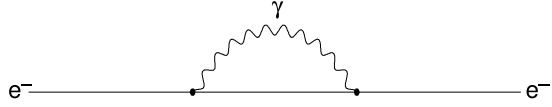}}
        \caption{The Coulomb self-energy of the electron.}
        \protect\label{fig:Coulomb}
\end{figure}

\begin{figure}[h]
        \centerline{\includegraphics[scale=0.5]{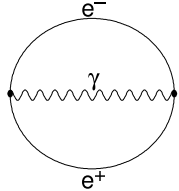}}
        \caption{The bubble diagram which shows the fluctuation of the vacuum.}
        \protect\label{fig:bubble}
\end{figure}

\begin{figure}[h]
        \centerline{\includegraphics[scale=0.45]{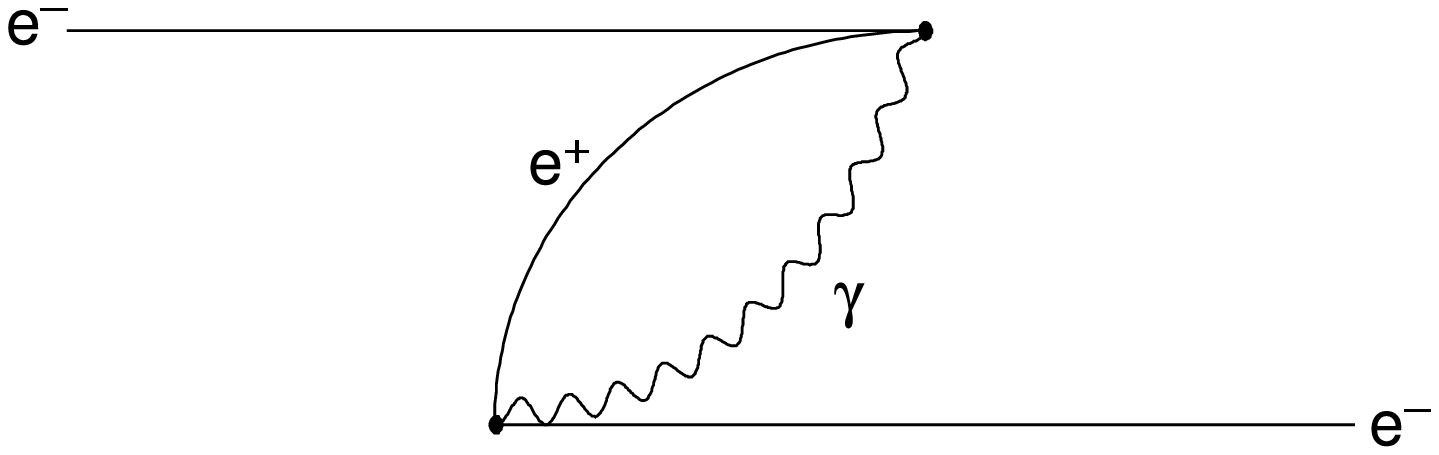}}
        \caption{Another contribution to the electron self-energy due to 
        the fluctuation of the vacuum.}
        \protect\label{fig:self2}
\end{figure}

The resolution to this problem came from the discovery of the
anti-particle of the electron, the positron, or in other words by
doubling the degrees of freedom in the theory.  The Coulomb
self-energy discussed above can be depicted by a diagram
Fig.~\ref{fig:Coulomb} where the electron emits the Coulomb field (a
virtual photon) which is absorbed later by the electron (the electron
``feels'' its own Coulomb field).\footnote{The diagrams
  Figs.~\ref{fig:Coulomb}, \ref{fig:self2} are not Feynman diagrams,
  but diagrams in the old-fashioned perturbation theory with different
  $T$-orderings shown as separate diagrams.  The Feynman diagram for
  the self-energy is the same as Fig.~\ref{fig:Coulomb}, but
  represents the {\it sum}\/ of Figs.~\ref{fig:Coulomb},
  \ref{fig:self2} and hence the linear divergence is already cancelled
  within it.  That is why we normally do not hear/read about linearly
  divergent self-energy diagrams in the context of field theory.} But
now that we know that the positron exists (thanks to Anderson back in
1932), and we also know that the world is quantum mechanical, one
should think about the fluctuation of the ``vacuum'' where the vacuum
produces a pair of an electron and a positron out of nothing together
with a photon, within the time allowed by the energy-time uncertainty
principle $\Delta t \sim \hbar/\Delta E \sim \hbar/(2 m_{e} c^{2})$
(Fig.~\ref{fig:bubble}).  This is a new phenomenon which didn't exist
in the classical electrodynamics, and modifies physics below the
distance scale $d \sim c \Delta t \sim \hbar c/(2 m_{e} c^{2}) =
200\times 10^{-13}$~cm.  Therefore, the classical electrodynamics
actually did have a finite applicability only down to this distance
scale, much earlier than $2.8 \times 10^{-13}$~cm as exhibited by the
problem of the fine cancellation above.  Given this vacuum fluctuation
process, one should also consider a process where the electron sitting
in the vacuum by chance annihilates with the positron and the photon
in the vacuum fluctuation, and the electron which used to be a part of
the fluctuation remains instead as a real electron
(Fig.~\ref{fig:self2}).  V.~Weisskopf \cite{Weisskopf} calculated this
contribution to the electron self-energy, and found that it is
negative and cancels the leading piece in the Coulomb self-energy
exactly:\footnote{An earlier paper by Weisskopf actually found two
  contributions to add up.  After Furry pointed out a sign mistake, he
  published an errata with no linear divergence.  I thank Howie Haber
  for letting me know.}
\begin{equation}
    \Delta E_{\rm pair} = - \frac{1}{4\pi \varepsilon_{0}}\frac{e^{2}}{r_{e}}.
    \label{eq:Epair}
\end{equation}
After the linearly divergent piece $1/r_{e}$ is canceled, the 
leading contribution in the $r_{e} \rightarrow 0$ limit is given by
\begin{equation}
    \Delta E = \Delta E_{\rm Coulomb} + \Delta E_{\rm pair}
    = \frac{3\alpha}{4\pi} m_{e} c^{2} \log \frac{\hbar}{m_{e} c r_{e}}.
    \label{eq:DeltaE}
\end{equation}
There are two important things to be said about this formula.  First, 
the correction $\Delta E$ is proportional to the electron mass and 
hence the total mass is proportional to the ``bare'' mass of the 
electron,
\begin{equation}
    (m_{e} c^{2})_{\it obs} = (m_{e}c^{2})_{\it bare}
    \left[ 1 + \frac{3\alpha}{4\pi} \log \frac{\hbar}{m_{e} c r_{e} }
    \right].
    \label{eq:self2}
\end{equation}
Therefore, we are talking about the ``percentage'' of the correction, 
rather than a huge additive constant.  Second, the correction depends 
only logarithmically on the ``size'' of the electron.  As a result, 
the correction is only a 9\% increase in the mass even for an electron 
as small as the Planck distance $r_{e} = 1/M_{Pl} = 1.6 \times 
10^{-33}$~cm.  

The fact that the correction is proportional to the ``bare'' mass is 
a consequence of a new symmetry present in the theory with the 
antiparticle (the positron): the chiral symmetry.  In the limit of 
the exact chiral symmetry, the electron is massless and the symmetry 
protects the electron from acquiring a mass from self-energy 
corrections.  The finite mass of the electron breaks the chiral 
symmetry explicitly, and because the self-energy correction should 
vanish in the chiral symmetric limit (zero mass electron), the 
correction is proportional to the electron mass.  Therefore, the 
doubling of the degrees of freedom and the cancellation of the power 
divergences lead to a sensible theory of electron applicable to very 
short distance scales.

\subsection{Hierarchy Problem}

In the Standard Model, the Higgs potential is given by
\begin{equation}
    V = m^{2} |H|^{2} + \lambda |H|^{4},
    \label{eq:V}
\end{equation}
where $v^{2} = \langle H \rangle^{2} = -m^{2}/2\lambda = (176~{\rm
GeV})^{2}$.  Because perturbative unitarity requires that $\lambda
\lesssim 1$, $-m^{2}$ is of the order of $(100~{\rm GeV})^{2}$. 
However, the mass squared parameter $m^{2}$ of the Higgs doublet
receives a quadratically divergent contribution from its self-energy
corrections.  For instance, the process where the Higgs doublets
splits into a pair of top quarks and come back to the Higgs boson
gives the self-energy correction
\begin{equation}
   \Delta m^{2}_{\rm top} = - 6 \frac{h_{t}^{2}}{4\pi^{2}} 
   \frac{1}{r_{H}^{2}},
    \label{eq:mu2top}
\end{equation}
where $r_{H}$ is the ``size'' of the Higgs boson, and $h_{t} \approx
1$ is the top quark Yukawa coupling.  Based on the same argument in
the previous section, this makes the Standard Model not applicable
below the distance scale of $10^{-17}$~cm.  This is the hierarchy
problem.  In other words, if we don't solve this problem, we can't
even talk about physics at much shorter distances without an excessive
fine-tuning in parameters.

It is worth pondering if the mother nature may fine-tune.  Now that
the cosmological constant appears to be fine-tuned at the level of
$10^{-120}$, should we be really worried about the fine-tuning of
$v^2/M_{Pl}^2 \approx 10^{-30}$ \cite{Arkani-Hamed:2004fb}?  In fact,
some people argued that the hierarchy exists because intelligent life
cannot exist otherwise \cite{Agrawal:1997gf}.  On the other hand, a
different way of varying the hierarchy does seem to support stellar
burning and life \cite{Harnik:2006vj}.  I don't get into this debate
here, but I'd like to just point out that a different fine-tuning
problem in cosmology, horizon and flatness problems, pointed to the
theory of inflation, which in turn appears to be empirically supported
by data.  I just hope that proper solutions will be found to both of
these fine-tuning problems and we will see their manifestations at the
relevant energy scale, namely TeV.  You have to be an optimist to work
on big problems.

\section{Examples of Physics Beyond the Standard Model}

Given various problems in the standard model discussed in the previous
section, especially the hierarchy problem, many possible directions of
physics beyond the standard model have been proposed.  I can review
only a few of them here given the spacetime constraint.  But I
especially emphasize the aspect of the models that leads to a (nearly)
stable neutral particle as a good dark matter candidate.

\subsection{Supersymmetry\label{sec:supersymmetry}}

The motivation for supersymmetry is to make the Standard Model
applicable to much shorter distances so that we can hope that the
answers to many of the puzzles in the Standard Model can be given by
physics at shorter distance scales \cite{motivation}.  In order to do
so, supersymmetry repeats what history did with the positron: doubling
the degrees of freedom with an explicitly broken new symmetry.  Then
the top quark would have a superpartner, the stop,\footnote{This is a
  terrible name, which was originally meant to be ``scalar top'' or
  ``supersymmetric top.''  Some other names are even worse: {\it
    sup}\/, {\it sstrange}\/, etc.  If supersymmetry will be
  discovered at LHC, we should seriously look for better names for the
  superparticles, maybe after the names of rich donors.}  whose loop
diagram gives another contribution to the Higgs boson self energy
\begin{equation}
   \Delta m^{2}_{\rm stop} = + 6 \frac{h_{t}^{2}}{4\pi^{2}} 
   \frac{1}{r_{H}^{2}}.
    \label{eq:mu2stop}
\end{equation}
The leading pieces in $1/r_{H}$ cancel between the top and stop 
contributions, and one obtains the correction to be
\begin{equation}
    \Delta m^{2}_{\rm top} + \Delta m^{2}_{\rm top}
    = -6\frac{h_{t}^{2}}{4\pi^{2}} (m_{\tilde{t}}^{2} - m_{t}^{2})
    \log \frac{1}{r_{H}^{2} m_{\tilde{t}}^{2}}.
    \label{eq:Deltamu2}
\end{equation}

One important difference from the positron case, however, is that the
mass of the stop, $m_{\tilde{t}}$, is unknown.  In order for the
$\Delta m^{2}$ to be of the same order of magnitude as the tree-level
value $m^{2} = -2\lambda v^{2}$, we need $m_{\tilde{t}}^{2}$ to be not
too far above the electroweak scale.  TeV stop mass is already a fine
tuning at the level of a percent.  Similar arguments apply to masses
of other superpartners that couple directly to the Higgs doublet.
This is the so-called naturalness constraint on the superparticle
masses (for more quantitative discussions, see papers
\cite{naturalness}).

Supersymmetry doubles the number of degrees of freedom in the standard
model.  For each fermion (quarks and leptons), you introduce a complex
scalar field (squarks and sleptons).  For each gauge boson, you
introduce gaugino, a partner Majorana fermion (a fermion field whose
anti-particle is itself).  I do not go into technical aspect of how to
write a supersymmetric quantum field theory; you should consult some
review articles \cite{Murayama:2000dw,Martin:1997ns}.

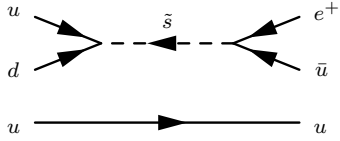
\begin{figure}[h]
  \centering
  \begin{fmffile}{pdecay}
    \begin{fmfgraph*}(100,40) 
      \fmfstraight
      \fmfleft{i3,i2,i1}
      \fmfright{o3,o2,o1}
      \fmf{fermion}{i1,v1}
      \fmflabel{$u$}{i1}
      \fmf{fermion}{i2,v1}
      \fmflabel{$d$}{i2}
      \fmf{scalar,label=$\tilde{s}$}{v2,v1}
      \fmf{fermion}{o1,v2}
      \fmflabel{$e^+$}{o1}
      \fmf{fermion}{o2,v2}
      \fmflabel{$\bar{u}$}{o2}
      \fmf{fermion}{i3,o3}
      \fmflabel{$u$}{i3}
      \fmflabel{$u$}{o3}
    \end{fmfgraph*}
  \end{fmffile}
  \caption{A possible Feynman diagram with supersymmetric particles
    that can lead to a too-rapid proton decay $p \rightarrow e^+
    \pi^0$.}\label{fig:pdecay}
\end{figure}

One important point related to dark matter is the proton longevity.
We know from experiments such as SuperKamiokande that proton is {\it
  very}\/ long lived (if not immortal).  The life time for the decay
mode $p \rightarrow e^+ \pi^0$ is longer than $1.6 \times
10^{33}$~years, at least twenty-three orders of magnitude longer than
the age of the universe!  On the other hand, if you write the most
general renormalizable theory with standard model particle content
consistent with supersymmetry, it allows for vertices such as
$\epsilon_{ijk} u^i d^j \tilde{s}^k$ and $e u^i \tilde{s}^*_i$ (here
$i,j,k$ are color indices).  Then one can draw a Feynman diagram like
one in Fig.~\ref{fig:pdecay}.  If the couplings are $O(1)$, and
superparticles around TeV, one finds the proton lifetime as short as
$\tau_p \sim m_{\tilde{s}}^4 / m_p^5 \sim 10^{-12}$~sec; {\it a little
  too short!}

Because of this embarrassment, we normally introduce a ${\mathbb Z}_2$
symmetry called ``$R$-parity'' defined by
\begin{equation}
  R_p = (-1)^{3B+L+2s} = (-1)^{\rm matter} R_{2\pi}
\end{equation}
where $s$ is the spin.  What it does is to flip the sign of all matter
fields (quarks and leptons) and perform $2\pi$ rotation of space at
the same time.  In effect, it assigns even parity to all particles in
the standard model, and odd parity to their superpartners.  Here is a
quick check.  For the quarks, $B=1/3$, $L=0$, and $s=1/2$, and we find
$R_p = +1$, while for squarks the difference lies in $s=0$ and hence
$R_p = -1$.  This symmetry forbids both of the bad vertices in
Fig.~\ref{fig:pdecay}.

Once the $R$-parity is imposed,\footnote{An obvious objection is that
  imposing $R$-parity appears {\it ad hoc.}\/ Fortunately there are
  several ways for it to emerge from a more fundamental theory.
  Because the $R$-parity is anomaly-free \cite{Ibanez:1991pr}, it may
  come out from string theory.  Or $R_p$ can arise as a subgroup of
  the $SO(10)$ grand unified gauge group because the matter belongs to
  the spinor representation and Higgs to vector, and hence $2\pi$
  rotation in the gauge group leads precisely to $(-1)^{\rm matter}$.
  It may also be an accidental symmetry due to other symmetries of the
  theory \cite{Murayama:1994tc,Carone:1996nd} so that it is slightly
  broken and dark matter may eventually decay.} there are no baryon-
and lepton-number violating interaction you can write down in a
renormalizable Lagrangian with the standard model particle content.
This way, the $R$-parity makes sure that proton is long lived.  Then
the lightest supersymmetric particle (LSP), with odd $R$-parity,
cannot decay because there are no other states with the same
$R$-parity with smaller mass it can decay into by definition.  In most
models it also turns out to be electrically neutral.  Then one can
talk about the possibility that the LSP is the dark matter of the
universe.

\subsection{Composite Higgs}

Another way the hierarchy problem may be solved is by making the Higgs
boson to actually have a finite size.  Then the correction in
Eq.~(\ref{eq:mu2top}) does not require tremendous fine-tuning as long
as the physical size of the Higgs boson is about $r_H \approx ({\rm
  TeV})^{-1} \approx 10^{-17}~\mbox{cm}$.  This is possible if the
Higgs boson is a composite object made of some elementary
constituents.

The original idea along this line is called technicolor (see reviews
\cite{technicolor,Hill:2002ap}), where a new strong gauge force binds
fermions and anti-fermions much like mesons in the real QCD.  Again
just like in QCD, fermion anti-fermion pair have a condensate $\langle
\bar{\psi} \psi \rangle \neq 0$ breaking chiral symmetry.  In
technicolor theories, this chiral symmetry breaking is nothing but the
breaking of the electroweak $SU(2) \times U(1)$ symmetry to the $U(1)$
QED subgroup.  Because the Higgs boson is heavy and strongly
interacting, it is expected to be too wide to be seen as a particle
state.

It is fair to say, however, that the technicolor models suffer from
various problems.  First of all, it is difficult to find a way of
generating sufficient masses for quarks and leptons, especially the
top quark, because you have to rely on higher dimension operators of
type $\bar{q} q \bar{\psi} \psi/\Lambda^2$.  The scale $\Lambda$ must
be low enough to generate $m_t$, while high enough to avoid excessive
flavor-changing neutral current.  In addition, there is tension with
precision electroweak observables.  These observables are precise
enough that they constrain heavy particles coupled to $Z$- and
$W$-bosons even though we cannot produce them directly.\footnote{It is
  curious that higher dimensional versions of technicolor models
  called Higgsless models \cite{Csaki:2003dt} do much better
  \cite{Csaki:2003zu}.  A supersymmetric version of technicolor also
  does better than the original technicolor \cite{Murayama:2003ag}.  }

Because of this issue, there are various other incarnations of
composite Higgs idea, which try to get a relatively light Higgs boson
as a bound state \cite{Kaplan:1983fs,Georgi:1984af}.  One of the
realistic models is called ``little Higgs''
\cite{Arkani-Hamed:2002qx,Arkani-Hamed:2002qy}.  Because of the
difficulty of achieving Higgs compositeness at the TeV scale, we are
better off putting off the compositeness scale to about 10~TeV to
avoid various phenomenological constraints.  Then you must wonder if
the problem with Eq.~(\ref{eq:mu2top}) comes back.  But there is a way
of protecting the scale of Higgs mass much lower than the
compositeness scale by using symmetries similar to the reason why a
pion is so much lighter than a proton.  If you are clever,
you can arrange the structure of symmetry such that it eliminates the
one-loop correction in Eq.~(\ref{eq:mu2top}) and the correction arises
only at the two-loop level.  Then the compositeness $\sim 10$~TeV is
not a problem.

Another attractive idea is to use extra dimensions to generate the
Higgs field from a gauge field, called ``Higgs-gauge unification''
\cite{Manton:1979kb,Hall:2001zb,vonGersdorff:2002as,Csaki:2002ur,Burdman:2002se}.
We know the mass of the gauge boson is forbidden by the gauge
invariance.  If the Higgs field is actually a gauge boson (spin one),
but if it is spinning in extra dimensions, we (as observers stuck in
four dimensions) perceive it not to spin.  Not only this gives us {\it
  raison d'\^etre}\/ of (apparently) spinless degrees of freedom, it
also provides protection for the Higgs mass and hence solves the
hierarchy problem.  The best implementation of this line of thinking
is probably the holographic Higgs model in
Refs.~\cite{Contino:2003ve,Agashe:2004rs} which involves the warped
extra dimension I will briefly discuss in the next section.  It should
also be said that many of the ideas mentioned here are closely related
to each other \cite{Cheng:2006ht}.

Similarly to the case of supersymmetry, people often introduce a
${\mathbb Z}_2$ symmetry to avoid certain phenomenological
embarrassments.  In little Higgs theories, tree-level exchange of new
particles tend to cause tension with precision electroweak
constraints.  Then the new states must be sufficiently heavy so that
the hierarchy problem is reintroduced.  By imposing ``$T$-parity,''
new particles can only appear in loops for low-energy processes and
the constraints can be easily avoided \cite{Cheng:2004yc}.  Then the
lightest $T$-odd particle (LTP) becomes a candidate for dark matter.
In technicolor models, the lightest technibaryon is stable (just like
proton in QCD) and a dark matter candidate \cite{Nussinov:1985xr}.

\subsection{Extra Dimensions}

The source of the hierarchy problem is our thinking that there {\it
  is}\/ physics at much shorter distances than $10^{-17}$~cm.  What if
there isn't?  What if physics {\it ends}\/ at $10^{-17}$~cm where
quantum field theory stops applicable and is taken over by something
more radical such as string theory?  Normally we associate the
ultimate limit of field theory with the Planck scale $d_{Pl} \approx
10^{-33}~\mbox{cm} = G_N^{1/2}$ where the gravity becomes as strong as
other forces and its quantum effects can no longer be ignored.  How
then can the quantum gravity effects enter at a much larger distance
scale such as $10^{-17}$~cm?

One way is to contemplate ``large'' extra dimensions of size $R$
\cite{ADD}.  Imagine there are $n$ extra dimensions in addition to our
three-dimensional space.  If you place two test masses at a distance
$r$ much shorter than $R$, the field lines of gravity spread out into
$3+n$ dimensions and the force decreases as the surface area
$r^{-(n+2)}$.  However, if the distance is longer than $R$, there is a
limit to which how much the field lines can spread because they are
squeezed within the size $R$.  Therefore, the force decreases only as
$r^{-2}$ for $r \gg R$, reproducing the usual inverse square law of
gravity.  It turns out that the inverse square law is tested only down
to 44~$\mu$m \cite{Kapner:2006si} (even though this is very
impressive!) and extra dimensions smaller than that are allowed
experimentally.

Matching two expressions for the gravitational force at $r=R$, we can
related the Newton's constant in $n+3$ dimensions to the Newton's
constant $G_N$ in three dimensions
\begin{equation}
  G_{n+3} \frac{1}{R^{n+2}} = G_N \frac{1}{R^2}, 
\end{equation}
and hence $G_{n+3} = G_N R^n$.  In the natural unit $\hbar=c=1$, the
mass scale of gravity is related to the Planck scale by $M^{n+2} R^n =
M_{Pl}^2$.  Even if the true energy scale of quantum gravity is at $M
\sim 1$~TeV, we may find an {\it apparent}\/ scale of gravity to be
$M_{Pl} \sim 10^{19}$~GeV.  Then the required size of extra dimensions
is
\begin{equation}
  R = \left\{
    \begin{array}{ll}
      10^{15}~\mbox{cm} & (n=1)\\
      10^{-1}~\mbox{cm} & (n=2)\\
      10^{-6}~\mbox{cm} & (n=3)\\
      10^{-12}~\mbox{cm} & (n=6)\\
    \end{array}
  \right. 
\end{equation}
Obviously the $n=1$ case is excluded because $R$ is even bigger than
1AU~$\approx 10^{13}$~cm.  The case $n=2$ is just excluded by the
small-scale gravity experiment, while $n\leq 3$ is completely
allowed.  

\begin{figure}[h]
  \centering
  \includegraphics[width=0.3\textwidth]{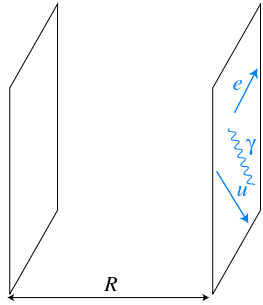}
  \caption{Large extra dimensions.  Even though the three-brane is
    drawn at the ends of extra dimensions, it does not have to be.}
  \label{fig:ADD}
\end{figure}

If we don't see the extra dimensions directly, what do they do to us?
Let us look at the case of just one extra dimension with periodic
boundary condition $y \rightarrow y + 2\pi R$.  Then all particles
have wave functions on the coordinate $y$ that satisfies $\psi(y+2\pi
R) = \psi(y)$.  They can of course depend on the usual
four-dimensional space time $x$, too.  One can expand it in Fourier modes
\begin{equation}
  \psi(x, y) = \sum_n \psi_n(x) e^{iny/R}.
\end{equation}
The momentum along the $y$ direction is $p_y = -i \partial_y = n/R$,
and the total energy of the particle is
\begin{equation}
  E = \sqrt{\vec{p}^2 + p_y^2}
  = \sqrt{\vec{p}^2 + \left( \frac{n}{R} \right)^2 }.
\end{equation}
Namely that you find a tower of particles of mass $m_n = n/R$, called
Kaluza--Klein states.  

Of course, the standard model is tested down to $10^{-17}$~cm, and we
have not found Kaluza--Klein excitation of electron, etc.  This is not
a problem if we are stuck on a three-dimensional sheet (three-{\it
  brane}) embedded inside the $n+3$ dimensional space.  The branes are
important objects in string theory and it is easy to get particles
with gauge interactions stuck on them.  The brane may be freely
floating inside extra dimensions or may be glued at singularities
({\it e.g.}\/, orbifold fixed points).  The simplest way to use large
extra dimensions is to assume that only gravity is spread out in extra
dimensions, while the standard model particles are all on a
three-brane.

Cosmology with large extra dimension is an iffy subject, however.  The
Kaluza--Klein excitation of gravitons can be produced in early
universe and the cosmology would be different from the standard
Friedmann univese (see, {\it e.g.}\/, \cite{Hall:1999mk}).  I~will not
get into this discussion here.

Instead of models with large extra dimensions, models with small extra dimensions of size 
$R\approx 10^{-17}$~cm~$\approx {\rm TeV}^{-1}$ are also
interesting,\footnote{Historically, unified theories and string theory
  assumed $R \approx d_{Pl} \approx 10^{-33}$~cm.  TeV-sized extra
  dimensions are {\it much}\/ larger than this, but I'm calling them
  ``small'' for the sake of distinction from the large extra
  dimensions.}  which allow for normal cosmology below TeV
temperatures.  This would also allow us standard model particles to
live in extra dimensions, too, because our Kaluza--Klein excitations
have been too heavy to be produced at accelerators so far.  There are
many versions of small extra dimensions.

One very popular version is warped extra dimension
\cite{Randall:1999ee}.  Instead of flat metric in the extra
dimensions, it sets up an exponential behavior.  It is something like
Planck scale varies from $10^{19}$~GeV to TeV as you go across the 5th
dimension.  Therefore, physics does {\it end}\/ at TeV if you on one
end of the 5th dimension, while it keeps going to $10^{19}$~GeV on the
other end.  The hierarchy problem may be solved if Higgs resides on
(or close to) the ``TeV brane.''  This set up attracted a lot of
attention because the bulk is actually a slice of anti-de Sitter space
which has nice features of preserving supersymmetry, leading to
AdS/CFT correspondence \cite{Maldacena:1997re}, etc.  It is also
possible to obtain quite naturally from string theory
\cite{Giddings:2001yu}.  In a grand unified model from warped extra
dimension, the proton longevity is an issue which is solved by a
${\mathbb Z}_3$ symmetry, and the lightest ${\mathbb Z}_3$-charged particle
(LZP) is a candidate for dark matter \cite{Agashe:2004ci}.

\begin{figure}[h]
  \centering
  \includegraphics[width=0.3\textwidth]{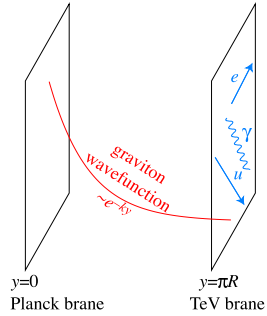}
  \caption{Warped extra dimension.  Even though the standard model
    particles are shown to be on the TeV brane, they may propagate in
    the bulk depending on the models. }
  \label{fig:RS}
\end{figure}

\begin{figure}[h]
  \centering
  \includegraphics[width=0.5\textwidth]{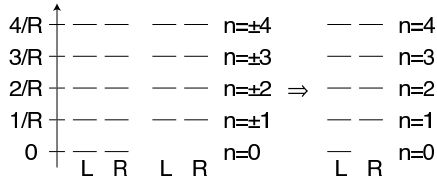}  
  \caption{The spectrum of fermions in the 5D bulk.  After orbifold
    identification in Fig.~\ref{fig:orbifold}, the spectrum is halved
    and one can obtain chiral fermions in 4D.}
  \label{fig:UED}
\end{figure}

It is also possible to have the ``flat'' extra dimension at the TeV
scale and put all the standard model particles in the 5D bulk, called
Universal Extra Dimension (UED) \cite{Appelquist:2000nn}.  It is tricky to
get chiral fermions in four dimensions if they are embedded in higher
dimensional space.  If you start out with five-dimensional Dirac
equation
\begin{equation}
  (i \gamma^\mu \partial_\mu + \gamma^5 \partial_y) \psi(x, y) = 0,
\end{equation}
the Fourier-mode expansion for the mode $\psi_n (x) e^{-iny/R}$ gives
\begin{equation}
  \left( i \gamma^\mu \partial_\mu - i \frac{n}{R} \gamma^5 \right)
  \psi_n (x) = 0.
\end{equation}
After a chiral rotation $\psi_n \rightarrow \psi_n e^{i\pi\gamma^5/2}
\psi_n$, the second term turns into the usual mass term without
$\gamma^5$.  The problem is that here are always two eigenvalues
$\gamma^5 = \pm 1$ and you find both left- and right-handed fermions
with the same quantum numbers.  Namely, you get Dirac fermion, not
Weyl fermion.  Then you don't get the standard model that
distinguishes left from right.  In terms of spectrum, what is on the
left of Fig.~\ref{fig:UED} is the spectrum because the Fourier modes
$n$ and $-n$ give the degenerate mass $n/R$ each of them with its own
Dirac fermion.  

\begin{figure}[h]
  \centering
  \includegraphics[width=0.3\textwidth]{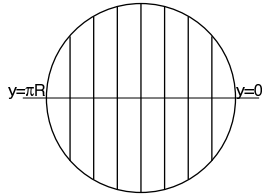}  
  \caption{The orbifold $S^1/{\mathbb Z}_2$. Points connected by the
    solid lines are identified.}
  \label{fig:orbifold}
\end{figure}

The trick to get chiral fermions is to use an orbifold
Fig.~\ref{fig:orbifold}.  Out of a circle $S^1$ ($y\in [-\pi R, \pi
R]$), you identify points $y$ and $-y$ to get a half-circle
$S^1/{\mathbb Z}_2$.  There are two special points, $y=0$ and $\pi R$,
that are identified only with themselves called ``fixed points.''  In
addition, we take the boundary condition that $\psi(y) = -\gamma^5
\psi(-y)$.  For $n\neq 0$, we use $\cos n y/R$ for $\gamma^5 = -1$ and
$\sin n y/R$ for $\gamma^5 = +1$, without the degeneracy between $n$
and $-n$.  For $n=0$, only $\gamma^5 = -1$ survives with the wave
function $\psi_0(y) = 1$.  This way, we keep only a half of the states
as shown in Fig.~\ref{fig:UED}, and we can get chiral fermions.  As a
consequence, we find the system to have a ${\mathbb Z}_2$ symmetry
under $y \rightarrow \pi R -y$, under which modes with even $n$ are
even and odd $n$ odd.  This ${\mathbb Z}_2$ symmetry is called KK
parity and the lightest KK state (LKP) becomes stable.  At the
tree-level, all first Kaluza--Klein states are degenerate $m_1 =
1/R$.\footnote{Here I've ignored possible complications due to brane
  operators and electroweak symmetry breaking.}  Radiative corrections
split their masses, and typically the first Kaluza--Klein excitation
of the $U(1)_Y$ boson is the LKP \cite{Cheng:2002iz}.  Because the
mass splittings are from the loop diagrams, they are small.  Similarly
to supersymmetry, there is a large number of new particles beyond the
standard model, namely Kaluza--Klein excitations.  Its collider
phenomenology very much resembles that of supersymmetry and it is not
trivial to tell them apart at the LHC (dubbed ``bosonic
supersymmetry'' \cite{Cheng:2002ab}).

\section{Evidence for Dark Matter}

Now we turn our attention to the problem of non-baryonic dark matter
in the universe.  Even though this is a sudden change in the topic,
you will see soon that it is connected to the discussions we had on
physics beyond the standard model.  We first review basics of
observational evidence for non-baryonic dark matter, and then discuss
how some of the interesting candidates are excluded.  It leads to a
paradigm that dark matter consists of unknown kind of elementary
particles.  By a simple dimensional analysis, we find that a weakly
coupled particle at the TeV-scale naturally gives the correct
abundance in the current universe.  We will take a look at a simple
example quite explicitly so that you can get a good feel on how it
works.  Then I will discuss more attractive dark matter candidates
that arise from various models of physics beyond the standard model I
discussed in the previous section.

The argument for the existence of ``dark matter,'' namely mass density
that is not luminous and cannot be seen in telescopes, is actually
very old.  Zwicky back in 1933 already reported the ``missing mass''
in Coma cluster of galaxies.  By studying the motion of galaxies in
the cluster and using the virial theorem (assuming of course that the
galactic motion is virialized) he determined the mass distribution in
the cluster and reported that a substantial fraction of mass is not
seen.  Since then, the case for dark matter has gotten stronger and
stronger and most of us regard its existence established by now.  I
refer to a nice review for details \cite{Raffelt:1997de} written back
in 1997, but I add some important updates since the review.

Arguably the most important one is the determination of cosmological
parameters by the power spectrum of CMB anisotropy.  In the fit to the
power-law flat $\Lambda$CDM model gives $\Omega_M h^2 =
0.127^{+0.007}_{-0.013}$ and $\Omega_B h^2 =
0.0223^{+0.0007}_{-0.0009}$ \cite{Spergel:2006hy}.  The point here is
that these two numbers are {\it different}\/.  Naively subtracting the
baryon component, and adding the errors by quadrature, I find
$(\Omega_M - \Omega_B)h^2 = 0.105^{+0.007}_{-0.013} \neq 0$ at a very
high precision.  This data alone says most of the matter component in
the universe is not atoms, something else.

Another important way to determine the baryon density of the universe
is based on Big-Bang Nucleosynthesis (BBN).  The baryon density is
consistent with what is obtained from the CMB power spectrum,
$\Omega_B h^2 = 0.0216^{+0.0020}_{-0.0021}$ from five best
measurements of deuterium abundance \cite{Cyburt:2004cq} using
hydrogen gas at high redshift (and hence believed to be primordial)
back-lit by quasars.  This agrees very well with the CMB result, even
though they refer to very different epochs: $T\sim 1$~MeV for BBN
while $T \sim 0.1$~eV for CMB.  This agreement gives us confidence
that we know $\Omega_B$ very well.


A novel technique to determine $\Omega_M$ uses large-scale structure,
namely the power spectrum in galaxy-galaxy correlation function.  As a
result of the acoustic oscillation in the baryon-photon fluid, the
power spectrum also shows the ``baryon oscillation'' which was
discovered only the last year \cite{Eisenstein:2005su}.  Without
relying on the CMB, they could determine $\Omega_M h^2 = 0.130 \pm
0.010$.  Again this is consistent with the CMB data, confirming the
need for non-baryonic dark matter.




I'd like to also mention a classic strong evidence for dark matter in
galaxies.  It comes from the study of rotation curves in spiral
galaxies.  The stars and gas rotate around the center of the galaxy.
For example, our solar system rotates in our Milky Way galaxy at the
speed of about 220~km/sec.  By using Kepler's law, the total mass
$M(r)$ within the radius $r$ and the rotation speed at this radius
$v(r)$ are related by
\begin{equation}
  v(r)^2 = G_N \frac{M(r)}{r}.
\end{equation}
Once the galaxy runs out of stars beyond a certain $r$, the rotation
speed is hence expected to decrease as $v(r) \propto r^{-1/2}$.  This
expectation is not supported by observation.

You can study spiral galaxies which happen to be ``edge-on.''  At the
outskirts of a galaxy, where you don't find any stars, there is cold
neutral hydrogen gas.  It turns out you can measure the rotation speed
of this cold gas.  A hydrogen atom has hyperfine splitting due to the
coupling of electron and proton spins, which corresponds to the famous
$\lambda =21$~cm line emission.  Even though the gas is cold, it is
embedded in the thermal bath of cosmic microwave background whose
temperature 2.7~K is {\it hot}\/ compared to the hyperfine excitation
$hc/k\lambda=0.069$~K.\footnote{If we had lived in a universe a
  hundred times larger, we would have lost this opportunity of
  studying dark matter content of the galaxies!}  Therefore the
hydrogen gas is populated in both hyperfine states and spontaneously
emits photons of wavelength 21~cm by the M1 transition.  This can be
detected by radio telescopes.  Because you are looking at the galaxy
edge-on, the rotation is either away or towards us, causing Doppler
shifts in the 21~cm line.  By measuring the amount of Doppler shifts,
you can determine the rotation speed.  Surprisingly, it was found that
the rotation speed stays constant well beyond the region where stars
cease to exist.

\begin{figure}[h]
  \centering
  \includegraphics[width=0.5\textwidth]{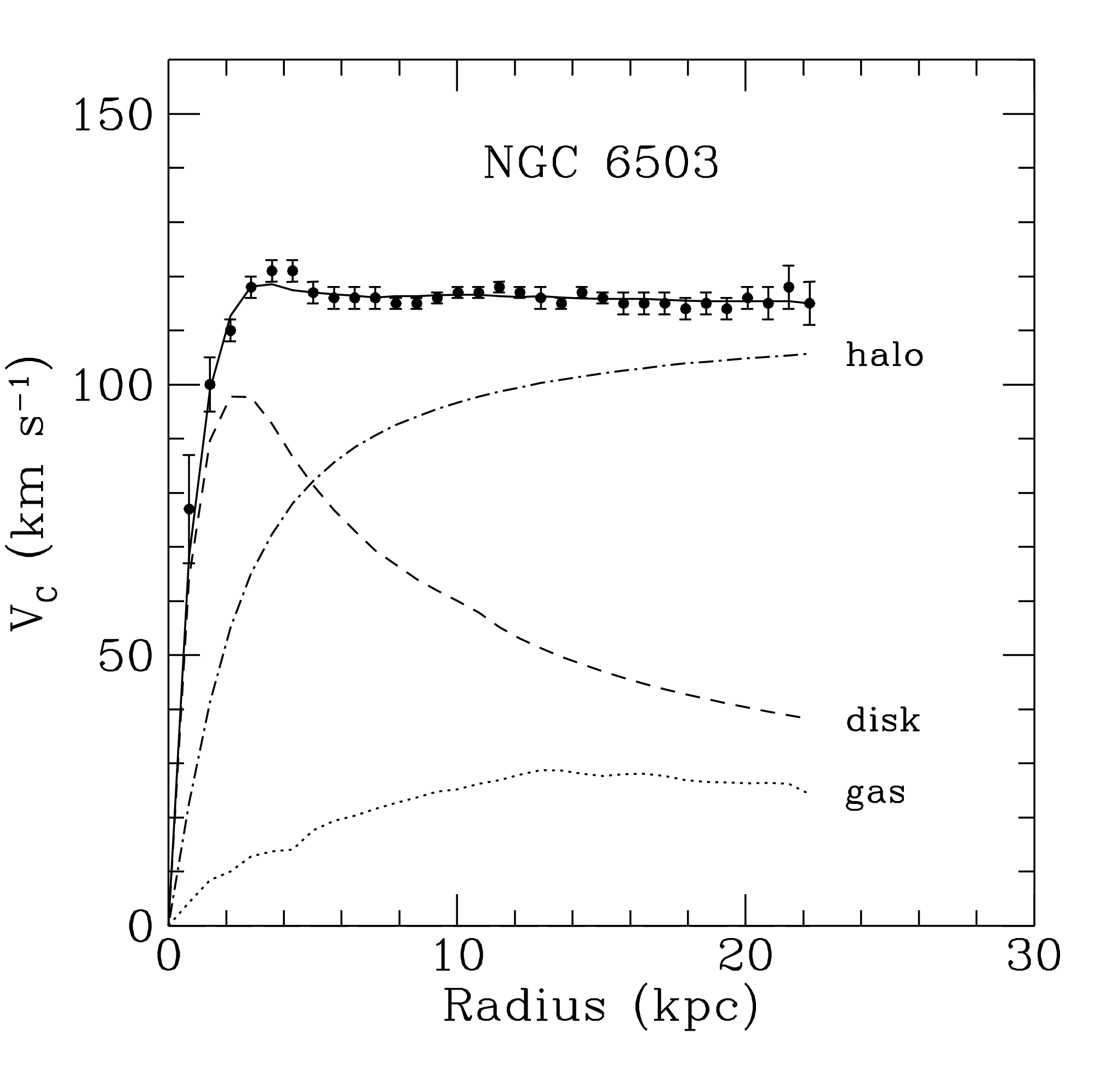}
  \caption{Rotation curve of a spiral galaxy \cite{Begeman:1991iy}.}
  \label{fig:rotationcurve}
\end{figure}

I mentioned this classic evidence because it really shows galaxies are
filled with dark matter.  This is an important point as we look for
signals of dark matter in our own galaxy.  It is not easy to determine
how much dark matter there is, however, because eventually the
hydrogen gas runs out and we do not know how far the flat rotation
curve extends.  Nonetheless, it shows the galaxy to be made up of a
nearly spherical ``halo'' of dark matter in which the disk is
embedded. 

\section{What Dark Matter Is Not}

We don't know what dark matter is, but we have learned quite a bit
recently what it is {\it not}\/.  I have already discussed that it is
{\it not}\/ ordinary atoms (baryons).  I~mention a few others of the
excluded possibilities.

\subsection{MACHOs}

The first candidate for dark matter that comes to mind is some kind of
astronomical objects, namely stars or planets, which are is too dark
to be seen.  People talked about ``Jupiters,'' ``brown dwarfs,'' etc.
In some sense, that would be the most conservative
hypothesis.\footnote{Somehow I can't call primordial black holes a
  ``conservative'' candidate without chuckling.} Because dark matter
is not made of ordinary atoms, such astronomical objects cannot be
ordinary stars either.  But one can still contemplate the possibility
that it is some kind of exotic objects, such as black holes.
Generically, one refers to MACHOs which stand for MAssive Compact Halo
Objects.

Black holes may be formed by some violent epochs in Big Bang
(primordial black holes or PBHs) \cite{Carr:1975qj} (see also
\cite{Hall:1989hr}).  If the entire horizon collapses into a black
hole, which is the biggest mass one can imagine consistent with
causality, for example in the course of a strongly first order phase
transition, the black hole mass would be
\begin{equation}
  M_{\rm PBH} \approx M_\odot \left(\frac{T}{\rm 100~MeV}\right)^{-2}
  \left( \frac{g_*}{10.75} \right)^{-1/2}.
\end{equation}
Therefore, there is no causal mechanism to produce PBHs much larger
than $10^3 M_\odot$ assuming that universe has been a normal radiation
dominated universe for $T \lesssim 3$~MeV to be compatible with
Big-Bang Nucleosynthesis.  Curiously, one finds $M_{\rm PBH} \approx
M_\odot$ if it formed at the QCD phase transition 
$T \approx 100$~MeV~\cite{Jedamzik:1996mr}.  On the other hand, PBHs cannot be too small
because otherwise they emit Hawking radiation of temperature $T=(8\pi
G_N M_{\rm PBH})^{-1}$ that would be visible.  The limit from diffuse
gamma ray background implies $M_{\rm PBH} \gtrsim 10^{-16}M_\odot$.

How do we look for such invisible objects?  Interestingly, it is not
impossible using the gravitational microlensing effects
\cite{Paczynski:1985jf}.  The idea is simple.  You keep monitoring
millions of stars in nearby satellite galaxies such as Large
Magellanic Cloud (LMC).  Meanwhile MACHOs are zooming around in the
halo of our galaxy at $v \approx 220$~km/s.  By pure chance, one of
them may pass very close along the line of sight towards one of the
stars you are monitoring.  Then the gravity would focus light around
the MACHO, effectively making the MACHO a lens.  You typically don't
have a resolution to observe distortion of the image or multiple
images, but the focusing of light makes the star appear temporarily
brighter.  This is called ``microlensing.'' By looking for such
microlensing events, you can infer the amount of MACHOs in our
galactic halo.

I've shown calculations on the deflection angle by the gravitational
lensing and the amplification in the brightness in the appendix.
(Just for fun, I've also added some discussions on the strong lensing
effects.)  The bottom line is that you may expect the microlensing
event at the rate of
\begin{equation}
  \mbox{rate} \approx 5 \times 10^{-6} \frac{1}{\rm year} \left(
    \frac{M_\odot}{M_{\rm MACHO}} \right)^{1/2}
\end{equation}
towards the LMC, with the duration of
\begin{equation}
  \mbox{duration} \approx 6 \times 10^6 \mbox{sec} \left( \frac{M_{\rm
        MACHO}}{M_\odot} \right)^{1/2} \left( \frac{\sqrt{d_1
        d_2}}{25~\mbox{kpc}} \right),
\end{equation}
where $d_1$ ($d_2$) is the distance between the MACHO and us (the
lensed star).  

\begin{figure}[h]
  \centering
  \includegraphics[width=0.5\textwidth]{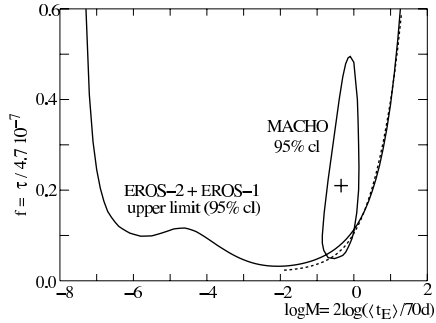}
  \caption{Limit on the halo fraction $f$ of MACHOs from the EROS
    collaboration \cite{Tisserand:2006zx}.  The spherical isothermal
  model of halo predicts the optical depth towards the LMC of $\tau =
  4.7 \times 10^{-7}$.  For more details, see the paper.}
  \label{fig:EROS}
\end{figure}

Two collaborations, the MACHO collaboration and the EROS
collaboration, have looked for microlensing events.  The basic
conclusion is that MACHOs of mass $10^{-7}$--30~$M_\odot$ cannot make
up 100\% of our galactic halo (Fig.~\ref{fig:EROS}).  See also
\cite{Alcock:1998fx,Tisserand:2006zx}.

Even though the possibility of MACHO dark matter may not be completely
closed, it now appears quite unlikely.  The main paradigm for the dark
matter of the universe has shifted from MACHOs to WIMPs. 

\subsection{Neutrinos}

Having discovered neutrinos have finite mass, it is also natural to
consider neutrinos to be dark matter candidate.  As a matter of fact,
neutrinos {\it are}\/ a component of dark matter, contributing
\begin{equation}
  \Omega_\nu h^2 = \frac{\sum_i m_{\nu_i}}{94~\mbox{eV}}.
\end{equation}
It is an attractive possibility if the particles which we already know
to exist could serve as the required non-baryonic dark matter.

However, as Sergio Pastor discussed in his lectures, neutrinos are not
good candidates for the bulk of dark matter for several reasons.
First, there is an upper limit on neutrino mass from laboratory
experiments (tritium beta decay) $m < 2$~eV \cite{Yao:2006px}.
Combined with the smallness of mass-squared differences $\Delta
m^2_\odot = 8 \times 10^{-5}$~eV$^2$ and $\Delta m^2_\oplus = 2.5
\times 10^{-3}$~eV$^2$, electron-volt scale neutrinos should be nearly
degenerate.  Then the maximum contribution to the matter density is
$\Omega_\nu h^2 < (3\times 2/94) < 0.064$.  This is not enough.

Second, even if the laboratory upper limit on the neutrino mass turned
out to be not correct, there is a famous Tremaine-Gunn argument
\cite{Tremaine:1979we}.  For the neutrinos to dominate the halo of
dwarf galaxies, you need to pack them so much that you would violate
Pauli exclusion principle.  To avoid this, you need to make neutrinos
quite massive $\gtrsim 500$~eV so that you don't need so many of them
\cite{Lin:1983vq}.  This obviously contradicts the requirement that
$\Omega_\nu < 1$. 

Third, neutrinos are so light that they are still moving at speed of
light (Hot Dark Matter) at the time when the structure started to
form, and erase structure at small scales.  Detailed study of large
scale structure shows such a hot component of dark matter must be
quite limited.  The precise limit depends on the exact method of
analyses.  A relatively conservative limit says $\sum_i m_{\nu_i} <
0.62$~eV \cite{Hannestad:2006mi} while a more aggressive limit goes
down to 0.17~eV \cite{Seljak:2006bg}.  Either way, neutrinos cannot
saturate what is needed for non-baryonic dark matter.

In fact, what we want is Cold Dark Matter, which is already
non-relativistic and slowly moving at the time of matter-radiation
equality $T \sim 1$~eV.  Naively a light (sub-electronvolt) particle
would not fit the bill.

A less conservative hypothesis may be to postulate that there is a new
heavy neutrino (4th generation).  This is a prototype for WIMPs that
will be discussed later.  It turns out, however, that the direct
detection experiments and the abundance do not have a compatible mass
range.  Namely the neutrinos are too strongly coupled to be the dark
matter!

\subsection{CHAMPs and SIMPs}

Even though people do not talk about it any more, it is worth
recalling that dark matter is unlikely be charged (CHAMP)
\cite{Dimopoulos:1989hk} or strongly interacting (SIMP)
\cite{Starkman:1990nj}.  I simply refer to papers that limit such
possibilities, from a multitude of search methods that include search
for anomalously heavy ``water'' molecule in the sea water, high-energy
neutrinos from the center of the Earth from annihilated SIMPs
accumulated there, collapsing neutron stars that accumulate CHAMPs,
etc.\footnote{I once got interested in the possibility that Jupiter
  is radiating heat more than it receives from the Sun because SIMPs
  are annihilating at its core \cite{Kawasaki:1991eu}.  It does not
  seem to explain heat from other Jovian planets, however, once
  empirical limits on SIMPs are taken into account.}

\section{WIMP Dark Matter}

WIMP, or Weakly Interactive Massive Particle, is the main current
paradigm for explaining dark matter of the universe.  With MACHOs
pretty much gone, it is indeed attractive to make a complete shift
from astronomical objects as heavy as $M_\odot \approx 10^{57}$~GeV to
``heavy'' elementary particles of mass $\sim 10^2$~GeV.  I will
discuss why this mass scale is particularly interesting.

\subsection{WIMP\label{sec:WIMP}}

The idea of WIMP is very simple.  It is a relatively heavy elementary
particle $\chi$ so that accelerator experiments so far did not have
enough energy to create them, namely $m_{\chi} \gtrsim 10^2$~GeV.  On
the other hand, the Big Bang did once have enough energy to make them.

Let us follow the history from when $T \gtrsim m_{\chi}$.  WIMPs were
created as much as any other particles.  Once the temperature dropped
below $m_{\chi}$, even the universe stopped creating them.  If they
are stable, whatever amount that was produced was there, and the only
way to get rid of them was to get them annihilating each other into
more mundane particles ({\it e.g.}\/, quarks, leptons, gauge bosons).
However, the universe expanded and there were fewer and fewer WIMPs in
a given volume, and at some point WIMPs stopped finding each other.
Then they could not annihilate any more and hence their numbers become
fixed (``freeze-out'').  This way, the universe could still be left
with a certain abundance of WIMPs.  This mechanism of getting dark
matter is called ``thermal relics.''

Let us make a simple estimate of the WIMP abundance.  In radiation
dominated universe, the expansion rate is given by
\begin{equation}
  H = \frac{\dot{a}}{a} = g_*^{1/2} \frac{T^2}{M_{Pl}} \left(
    \frac{\pi^2}{90} \right)^{1/2},
\end{equation}
where $M_{Pl} = 1/\sqrt{8\pi G_N} = 2.4 \times 10^{18}$~GeV is the
reduced Planck scale.  For simple estimates, we regard
$(\pi^2/90)^{1/2} = 0.33 \approx 1$ and ignore many other factors of
$O(1)$.  Hence, $H \simeq g_*^{1/2} T^2/M_{Pl}$.  The entropy density
is correspondingly
\begin{equation}
  s = g_* T^3 \left( \frac{2\pi^2}{45} \right)^{1/2} \simeq g_* T^3. 
\end{equation}
Given the thermally averaged annihilation cross section $\langle
\sigma_{\it ann} v\rangle$, and the number density of WIMPs
$n_{\chi}$, the annihilation rate of a WIMP is
\begin{equation}
  \Gamma = \langle \sigma_{\it ann} v\rangle n_{\chi}.
\end{equation}
The annihilation stops at the ``freeze-out temperature'' $T_f$ when
$\Gamma \simeq H$, and hence
\begin{equation}
  n_\chi (T_f) \simeq g_*^{1/2} \frac{T_f^2}{\langle \sigma_{\it ann} v\rangle
    M_{Pl}} \ .
\end{equation}
The yield of WIMPs is defined by $Y_\chi = n_\chi / s$.  This is a
convenient quantity because it is conserved by the expansion of the
universe as long as the expansion is adiabatic, {\it i.e.}\/, no new
source of heat.  This is due to the conservation of both the total
entropy and total number of particles and their densities both scale
as $1/a^3$.  The estimate of the yield is
\begin{equation}
  Y_\chi \sim  g_*^{-1/2} \frac{1}{\langle \sigma_{\it ann} v\rangle T_f
    M_{Pl}} 
  = g_*^{-1/2} \frac{x_f}{\langle \sigma_{\it ann} v\rangle m_\chi
    M_{Pl}} \ .
\end{equation}
Here, we defined $T_f = m_\chi/x_f$.  We will see later from more
detailed calculations that $x_f \sim 20$.  The abundance in the
current universe is calculated using the yield and the current entropy
density, divided by the current critical density,
\begin{equation}
  \Omega_\chi = m_\chi \frac{n_\chi}{s} \frac{s_0}{\rho_c}
  \sim g_*^{-1/2} \frac{x_f}{\langle \sigma_{\it ann} v\rangle M_{Pl}}
  \frac{s_0}{\rho_c}\ .
\end{equation}
We use $s_0 = 2890~\mbox{cm}^{-3}$ and $\rho_c = 1.05 \times 10^{-5}
h^2~\mbox{GeV cm}^{-3}$, where the current Hubble constant is $H_0 =
100h~\mbox{km/sec/Mpc}$ with $h \approx 0.65$.  In order of obtain
$\Omega_\chi h^2 \sim 0.12$, we need
\begin{equation}
  \langle \sigma_{\it ann} v \rangle
  \sim g_*^{-1/2} x_f \frac{1.12 \times
    10^{-10}~\mbox{GeV}^{-2}}{\Omega_\chi h^2} \sim 10^{-9}~\mbox{GeV}^{-2}.
\end{equation}
Recall a typical annihilation cross section of a particle of mass
$m_\chi$ by a relatively weak interaction of electromagnetic strength
({\it e.g.}\/, $e^+ e^- \rightarrow \gamma\gamma$) is
\begin{equation}
  \sigma_{\it ann} v \sim \frac{\pi \alpha^2}{m_\chi^2}\ .
\end{equation}
To obtain the correct abundance, what we need is
\begin{equation}
  m_\chi \sim 300~\mbox{GeV}.
\end{equation}
This is a {\it very}\/ interesting result.  Namely, the correct
abundance of thermal relics is obtained for a particle mass just
beyond the past accelerator limits and where we expect new particles
to exist because of the considerations of electroweak symmetry
breaking and the hierarchy problem.  In other words, it is exactly
the right mass scale for a new particle!

In the next few sections, we will firm up this naive estimate by
solving the Boltzmann equations numerically.  We will also study a
concrete model of a new particle for dark matter candidate and work
out its annihilation cross section.  In addition, we will see if we
have a chance of ``seeing'' the dark matter particle in our galactic
halo, or making it in future accelerator experiments.  Then we will
generalize the discussions to more theoretically attractive models of
physics beyond the standard model.

\subsection{Boltzmann Equation}

You have already seen Boltzmann equation in lectures by Sabino
Matarrese and I don't repeat its derivations.  We assume kinetic
equilibrium, namely that each particle species has the Boltzmann
distribution in the momentum space except for the overall
normalization that is given by its number density.  Considering the
process of $\chi_1 \chi_2 \leftrightarrow \chi_3 \chi_4$, where
$\chi_i$ refers to a certain elementary particle, the Boltzmann
equation for the number density $n_1$ for the particle $\chi_1$ is
\begin{equation}
  a^{-3} \frac{d (n_1 a^3)}{dt}
  = \langle \sigma v\rangle n_1^0 n_2^0
  \left( \frac{n_3 n_4}{n_3^0 n_4^0} - \frac{n_1 n_2}{n_1^0 n_2^0} \right).
\end{equation}
Here, $\sigma v$ is the cross section common for the process $\chi_1
\chi_2 \rightarrow \chi_3 \chi_4$ and its inverse process $\chi_3
\chi_4 \rightarrow \chi_1 \chi_2$ assuming the time reversal
invariance.  The number densities with superscript $^{0}$ refer to
those in the thermal equilibrium.

In the case of our interest, $\chi_{3,4}$ are ``mundane'' light
(relativistic) particles in the thermal bath, and hence $n_{3,4} =
n_{3,4}^0$.  In addition, we consider the annihilation $\chi \chi
\leftrightarrow (\mbox{mundane})^2$, and hence $n_1 = n_2$.  The
Boltzmann equation simplifies drastically to
\begin{equation}
  a^{-3} \frac{d n_\chi a^3}{dt}
  = \langle \sigma_{\it ann} v\rangle [(n_\chi^0)^2 - (n_\chi)^2].
\end{equation}
This time we pay careful attention to all numerical factors.  We use
\begin{eqnarray}
  Y &=& \frac{n_\chi}{s}, \\
  s &=& g_* T^3 \left( \frac{2\pi^2}{45} \right)^{1/2}, \\
  H^2 &=& \frac{8\pi}{3} G_N g_* \frac{\pi^2}{30} T^4
  = g_* \frac{\pi^2}{90} \frac{T^4}{M_{Pl}^2}, \\
  x &=& \frac{m_\chi}{T}\ .
\end{eqnarray}

Even though we start out at temperatures $T > m_\chi$ when $\chi$ are
relativistic, eventually the temperature drops below $m_\chi$ and we
can use non-relativistic approximations.  Then the equilibrium number
density can be worked out easily as
\begin{eqnarray}
  n_\chi^0 &=& \int \frac{d^3 p}{(2\pi)^3} e^{-E/T} \qquad\qquad
  \left( E = m_\chi + \frac{\vec{p}^2}{2m_\chi} \right) \nonumber \\
  &=& e^{-m_\chi/T} \left( \frac{m_\chi T}{2\pi} \right)^{3/2}
  = e^{-x} \frac{m_\chi^3}{(2\pi x)^{3/2}}\ .
\end{eqnarray}
Therefore
\begin{equation}
  Y_0 = \frac{n_\chi^0}{s} = \frac{1}{g_*} \frac{45}{2\pi^2} \left(
    \frac{x}{2\pi} \right)^{3/2} e^{-x}
  = 0.145 x^{3/2} e^{-x}.
\end{equation}

Changing the variables from $n_\chi$ to $Y$ and $t$ to $x$, the
Boltzmann equation becomes
\begin{equation}
  \frac{d Y}{d x} = - \frac{1}{x^2}
  \frac{s(m_\chi)}{H(m_\chi)} \langle \sigma_{\it ann} v\rangle
  (Y^2 - Y_0^2).
  \label{eq:Y}
\end{equation}
Here, we used $s(T) = s(m_\chi)/x^3$ and 
\begin{equation}
  dt = -\frac{1}{H(T)} \frac{dT}{T} =
  -\frac{m_\chi^2}{H(m_\chi) T^3} dT = \frac{1}{H(m_\chi)} x dx  .
\end{equation}
It is useful to work out
\begin{equation}
  \frac{s(m_\chi)}{H(m_\chi)}
  = \frac{2\pi^2}{45} \left( \frac{90}{\pi^2} \right)^{1/2} g_*^{1/2}
  m_\chi M_{Pl}
  = 1.32 g_*^{1/2} m_\chi M_{Pl}.
\end{equation}
Note that the annihilation cross section $\langle \sigma_{\it ann}
v\rangle$ is insensitive to the temperature once the particle is
non-relativistic $T \ll m_\chi$.\footnote{This statement assumes that
  the annihilation is in the $S$-wave.  If it is in the $l$-wave,
  $\langle \sigma_{\it ann} v \rangle \propto v^{2l} \propto x^{-l}$. }
Therefore the whole combination $\frac{s(m_\chi)}{H(m_\chi)} \langle
\sigma_{\it ann} v\rangle$ is just a dimensionless number.  The only
complication is that $Y_0$ has a strong dependence on $x$.  We can
further simplify the equation by introducing the quantity
\begin{equation}
  y = \frac{s(m_\chi)}{H(m_\chi)} \langle \sigma_{\it ann} v\rangle Y.
\end{equation}
We obtain
\begin{equation}
  \frac{dy}{dx} = - \frac{1}{x^2} (y^2 - y_0^2), 
  \label{eq:y}
\end{equation}
with
\begin{equation}
  y_0 
  = 0.192 g_*^{-1/2} M_{Pl} m_\chi \langle \sigma_{\it ann} v\rangle x^{3/2} e^{-x}.
\end{equation}

\subsection{Analytic Approximation}

Here is a simple analytic approximation to solve Eq.~(\ref{eq:y}).  We
assume $Y$ tracks $Y_0$ for $x<x_f$.  On the other hand, we assume $y
\gg y_0$ for $x > x_f$ because $y_0$ drops exponentially as $e^{-x}$.
Of course this approximation has a discontinuity at $x=x_f$, but the
transition between these two extreme assumptions is so quick that it
turns out to be a reasonable approximation.  Then we can analytically
solve the equation for $x > x_f$ and we find
\begin{equation}
  \frac{1}{y(\infty)} - \frac{1}{y(x_f)}
  = \frac{1}{x_f}\ .
\end{equation}
Since $y(\infty) \ll y(x_f)$, we obtain the simple estimate
\begin{equation}
  y(\infty) = x_f.
\end{equation}
Given this result, we can estimate $x_f$ as the point where $y_0(x)$
drops down approximately to $x_f$,
\begin{equation}
  0.192 g_*^{-1/2} M_{Pl} m_\chi \langle \sigma_{\it ann} v\rangle
  x_f^{3/2} e^{-x_f} \approx x_f,
\end{equation}
and hence
\begin{eqnarray}
  x_f &\approx& \ln \left( \frac{0.192 m_\chi M_{Pl} \langle \sigma_{\it ann}
      v\rangle x_f^{1/2}}{g_*^{1/2}} \right)\nonumber \\
  &\approx& 24 + \ln \frac{m_\chi}{100~\mbox{GeV}} 
  + \ln \frac{\langle \sigma_{\it
      ann}v\rangle}{10^{-9}~\mbox{GeV}^{-2}}
  - \frac{1}{2} \ln \frac{g_*}{100}.
  \label{eq:xf}
\end{eqnarray}

\subsection{Numerical Integration}

\begin{figure}[h]
  \centering
  \includegraphics[width=0.5\textwidth]{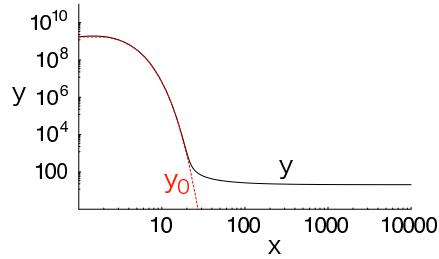}
  \caption{Numerical solution to the Boltzmann equation
    Eq.~(\ref{eq:y}) for $m=100$~GeV, $g_* = 100$, $\langle
    \sigma_{\it ann} v \rangle = 10^{-9}$~GeV$^{-2}$.  Superimposed
    is the equilibrium value $y_0$.} 
  \label{fig:Boltzmann}
\end{figure}

I've gone through numerical integration of the Boltzmann equation
Eq.~(\ref{eq:y}).  Fig.~\ref{fig:Boltzmann} shows the $x$-evolution of
$y$.  You can see that it traces the equilibrium value very well early
on, but after $x$ of about 20, it starts to deviate significantly and
eventually asymptotes to a constant.  This is exactly the behavior we
expected in the analytic approximation studied in the previous
section.

\begin{figure}[h]
  \centering
  \includegraphics[width=0.5\textwidth]{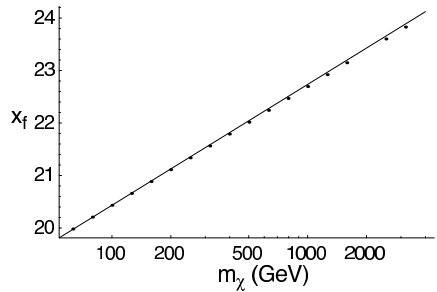}
  \caption{$x_f$ values as a function of $m_\chi$, for $g_* = 100$,
    $\langle \sigma_{\it ann} v \rangle = 10^{-9}$~GeV$^{-2}$.  The
    dots are the results of numerical integrations, while the solid
    line is just $\ln m_\chi$ with an offset so that $x_f = 20.43$ for
    $m_\chi = 100$~GeV. }
  \label{fig:xf}
\end{figure}

Fig.~\ref{fig:xf} shows the asymptotic values $y(\infty)$ which we
call $x_f$.  I understand this is a confusing notation, but we have to
define the ``freeze-out'' in some way, and the analytic estimate in
the previous section suggest that the asymptotic value $y(\infty)$ is
nothing but the freeze-out value $x_f$.  This is the result that
enters the final estimate of the abundance and is hence the only
number we need in the end anyway.  It does not exactly agree with the
estimate in the previous section, but does very well once I changed
the offset in Eq.~(\ref{eq:xf}) from 24 to 20.43.  Logarithmic
dependence on $m_\chi$ is verified beautifully.

Putting everything back together,
\begin{equation}
  \rho_\chi = m_\chi n_\chi
  = m_\chi Y s
  = m_\chi \frac{H(m_\chi)}{s(m_\chi)} \frac{x_f}{\langle \sigma_{\it
      ann} v\rangle} s
\end{equation}
As before, we use $s_0 = 2890~\mbox{cm}^{-3}$ and $\rho_c = 1.05
\times 10^{-5} h^2~\mbox{GeV cm}^{-3}$, where the current Hubble
constant is $H_0 = 100h~\mbox{km/sec/Mpc}$ with $h \approx 0.65$.  To
obtain $\Omega_M h^2 = 0.12$, we find $\langle \sigma_{\it
  ann}v\rangle = 1.6 \times 10^{-9}$~GeV$^{-2}$, confirming the simple
estimate in Section~\ref{sec:WIMP}.

\subsection{The New Minimal Standard Model}

Now we would like to apply our calculations to a specific model,
called the New Minimal Standard Model \cite{Davoudiasl:2004be}.  This
is the model that can account for the empirical facts listed in
Section~\ref{sec:empirical} with the minimal particle content if you
do not pay {\it any}\/ attention to the theoretical issues mentioned
in Section~\ref{sec:philosophical}.  It accomplishes this by adding
only four new particles to the standard model;\footnote{The other
  three are the inflaton and two right-handed neutrinos.} very minimal
indeed!  The dark matter in this model is a real scalar field $S$ with
an odd ${\mathbb Z}_2$ parity $S \rightarrow -S$, and its most general
renormalizable Lagrangian that should be added to the Standard Model
Eq.~(\ref{eq:LSM}) is
\begin{equation}
  \label{eq:S}
  {\cal L}_S =
  \frac{1}{2} \partial_\mu S \partial^\mu S - \frac{1}{2} m_S^2 S^2
  - \frac{k}{2} |H|^2 S^2 - \frac{h}{4!} S^4.
\end{equation}
The scalar field $S$ is the only field odd under ${\mathbb Z}_2$, and hence the
$S$ boson is stable.  Because of the analysis in the previous
sections, we know that if $m_S$ is at the electroweak scale, it may be
a viable dark matter candidate as a thermal relic.  This is a model
with only three parameters, $m_S$, $k$, and $h$, and actually the last
one is not relevant to the study of dark matter phenomenology.
Therefore this is a very predictive model where one can work it out
very explicitly and easily.

\begin{figure}[h]
  \centering
  \vspace{.5cm}
  \begin{fmffile}{SS1}
    \begin{fmfgraph*}(100,40) 
      \fmfleft{i1,i2}
      \fmfright{o1,o2}
      \fmf{dashes}{i1,v1,i2}
      \fmflabel{$S$}{i1}
      \fmflabel{$S$}{i2}
      \fmfv{label=$-ikv$}{v1}
      \fmf{dashes,label=$h$}{v1,v2}
      \fmf{fermion}{o1,v2,o2}
      \fmflabel{$\bar{f}$}{o1}
      \fmflabel{$f$}{o2}
      \fmfv{label=$-i\frac{m_f}{v}$}{v2}
    \end{fmfgraph*}
  \end{fmffile}
  \begin{fmffile}{SS2}
    \begin{fmfgraph*}(100,40) 
      \fmfleft{i1,i2}
      \fmfright{o1,o2}
      \fmf{dashes}{i1,v1,i2}
      \fmflabel{$S$}{i1}
      \fmflabel{$S$}{i2}
      \fmfv{label=$-ikv$}{v1}
      \fmf{dashes,label=$h$}{v1,v2}
      \fmf{photon}{o1,v2,o2}
      \fmflabel{$V=W,Z$}{o1}
      \fmflabel{$V=W,Z$}{o2}
      \fmfv{label=$ig_V m_V g^{\mu\nu}$}{v2}
    \end{fmfgraph*}
  \end{fmffile}
  \\ \vspace{20\lineskip}
  \begin{fmffile}{SS3}
    \begin{fmfgraph*}(100,40) 
      \fmfleft{i1,i2}
      \fmfright{o1,o2}
      \fmf{dashes}{i1,v1,i2}
      \fmflabel{$S$}{i1}
      \fmflabel{$S$}{i2}
      \fmfv{label=$-ikv$}{v1}
      \fmf{dashes,label=$h$}{v1,v2}
      \fmf{dashes}{o1,v2,o2}
      \fmflabel{$h$}{o1}
      \fmflabel{$h$}{o2}
      \fmfv{label=$-i \frac{3m_h^2}{v}$}{v2}
    \end{fmfgraph*}
  \end{fmffile}
  \begin{fmffile}{SS4}
    \begin{fmfgraph*}(100,40) 
      \fmfleft{i1,i2}
      \fmfright{o1,o2}
      \fmf{dashes}{i1,v1}
      \fmf{dashes,label=$S$}{v1,v2}
      \fmf{dashes}{v2,i2}
      \fmflabel{$S$}{i1}
      \fmflabel{$S$}{i2}
      \fmfv{label=$-ikv$}{v1}
      \fmfv{label=$-ikv$}{v2}
      \fmf{dashes}{o1,v1}
      \fmf{dashes}{o2,v2}
      \fmflabel{$h$}{o1}
      \fmflabel{$h$}{o2}
    \end{fmfgraph*}
  \end{fmffile}
  \begin{fmffile}{SS5}
    \begin{fmfgraph*}(100,40) 
      \fmfleft{i1,i2}
      \fmfright{o1,o2}
      \fmf{dashes}{i1,v1,i2}
      \fmf{dashes}{o1,v1,o2}
      \fmflabel{$S$}{i1}
      \fmflabel{$S$}{i2}
      \fmfv{label=$-ik$}{v1}
      \fmflabel{$h$}{o1}
      \fmflabel{$h$}{o2}
    \end{fmfgraph*}
  \end{fmffile}
  \vspace{.5cm}
  \caption{Feynman diagrams for the annihilation of $S$ scalars.  The
    final states in the first diagram can be any of the quark or
    lepton pairs $f\bar{f}$.}\label{fig:SS}
\end{figure}
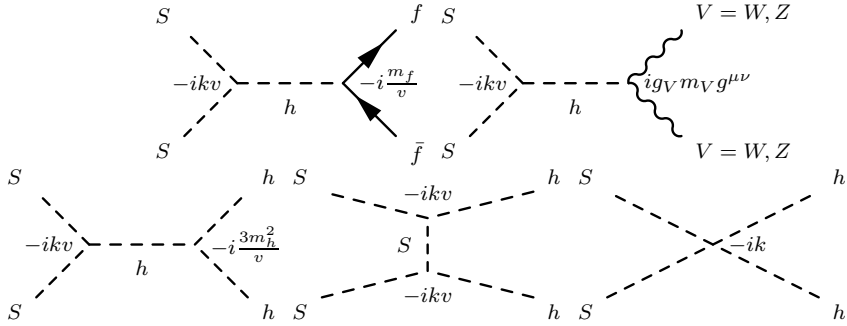

\begin{figure}
  \centering
  \includegraphics[width=0.5\textwidth]{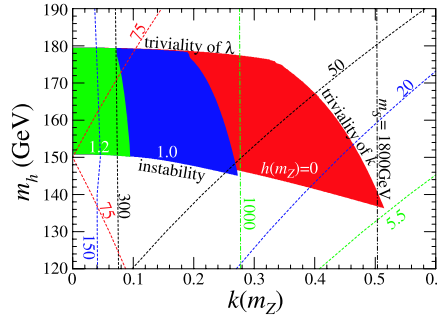}
  \caption{The region of the NMSM parameter space $(k(m_Z), m_h)$ that
    satisfies the stability and triviality bounds, for $h(m_Z)=0$,
    1.0, and 1.2.  Also the preferred values from the cosmic abundance
    $\Omega_S h^2 = 0.11$ are shown for various $m_S$.  Taken from
    \protect\cite{Davoudiasl:2004be}.}
  \label{fig:params}
\end{figure}

To calculate the dark matter abundance, what we need to know is the
annihilation cross section of the scalar boson $S$.  This was studied
first in \cite{McDonald:1993ex} and later in \cite{Burgess:2000yq},
but the third diagram was missing.  In addition, there is a
theoretical bounds on the size of couplings $k$ and $h$ so that they
would stay perturbative up to high scales ({\it e.g.}\/, Planck
scale).  The cosmic abundance is determined by $m_S$ and $k$ in
addition to $m_h$.  Therefore on the $(k,m_h)$ plane, the correct
cosmic abundance determines what $m_S$ should be.  This is shown in
Fig.~\ref{fig:params}.  You can see that for a very wide range $m_S
\simeq 5.5$~GeV--1.8~TeV, the correct cosmic abundance can be obtained
within the theoretically allowed parameter space.  For heavy $m_S \gg
m_h$, the cross section goes like $k^2/m_S^2$ and is independent of
$m_h$.  This is why the $m_S$ contours are approximately straight
vertically.  For light $m_S \ll m_h$, the cross section goes like $k^2
m_S^2/m_h^4$.  This is why the $m_S$ contours approximately have a
fixed $k/m_h^2$ ratio.  Note that when $m_S \simeq m_h/2$, the first
two diagrams can hit the Higgs pole and the cross section can be very
big even for small $k$.  This resonance effect is seen in
Fig.~\ref{fig:params} where $m_S = 75$~GeV line reaches almost $k=0$
for $m_h = 150$~GeV.

You may wonder why I am talking about $S$ as light as 5.5~GeV.
Shouldn't we have seen it already in accelerator experiments?
Actually, no.  The only interaction the $S$ boson has is with the
Higgs boson which we are yet to see.  Therefore, we could not have
produced the $S$ boson unless we had produced the Higgs boson.  That
is why even such a light $S$ boson does not contradict data.  In other
words, it wouldn't be easy to find this particle in accelerator
experiments.

\subsection{Direct Detection Experiments}

How do we know if dark matter is indeed in the form of WIMP candidate
you like?  One thing we'd love to see is the direct detection of
WIMPs.  The idea is very simple.  You place a very sensitive device in
a quite location.  WIMPs are supposed to be flying around in the halo
of our galaxy with the typical speed of $\sim 220$~km/s~$\sim
10^{-3}c$.  Because they are only very weakly interacting, they can go
through walls, rocks, even the entire Earth with little trouble, just
like neutrinos.  For a mass of $m_\chi \sim 100$~GeV, its typical
kinetic energy is $E_{\rm kin} = \frac{1}{2} m_\chi v^2 \sim 50$~keV.
If the WIMP (ever) scatters off an atomic nucleus, the energy deposit
is only (at most) of this order of magnitude.  It is a {\it tiny}\/
energy deposit that is very difficult to pick out against background
from natural radioactivity (typically MeV energies).  Therefore you
have to make the device very clean, and also place it deep underground
to be shielded from the cosmic-ray induced backgrounds, most
importantly neutrons ejected from the rocks by cosmic-ray muons.  One
you've done all this, what you do is to wait to see this little
``kick'' in your detector.

Let us do an order of magnitude estimate.  The local halo density is
estimated to be about $\rho_\chi^{\rm halo} \approx 0.3$~GeV/cm$^3$.
The number density of WIMPs is $n_\chi = \rho_\chi^{\rm halo}/m_\chi$.
The flux of WIMPs is roughly $v n_\chi \approx 10^{-3} c n_\chi$.  The
elastic cross section of WIMP on neutron or proton may be
spin-independent or spin-dependent.  In the spin-independent case, the
amplitude of the WIMP-nucleus cross section goes as $A$ (mass number)
and hence the cross section on the nucleus $\sigma_A$ goes as $A^2$.
Of course the detailed scaling is model-dependent, but in most
phenomenological analyses (and also analyses of data) we assume
$\sigma_A = A^2 \sigma_p$.  Let us also assume $^{56}$Ge as the
detector material so that $A=56$.  Then the expected event rate is
\begin{equation}
  R = n_\chi^{\rm halo} v \frac{m_{\rm target}}{m_A} \sigma_A
  \approx \frac{10}{\rm year}
  \frac{100~\mbox{GeV}}{m_\chi} \frac{m_{\rm target}}{100~\mbox{kg}}
  \frac{A}{56} \frac{\sigma_p}{10^{-42}~\mbox{cm}^2} \ .
\end{equation}
To prepare a very sensitive device as big as 100~kg and make it very
clean is a big job.  You can see that your wait may be {\it long}\/.

Now back to the New Minimal Standard Model.  The scattering of the $S$
boson off a proton comes from the $t$-channel Higgs boson exchange.
The coupling of the Higgs boson to the nucleon is estimated by the
famous argument \cite{Shifman:1978zn} using the conformal anomaly.
The mass of the proton is proportional to the QCD scale $m_p \propto
\Lambda_{QCD}$ ($m_u, m_d, m_s$ are ignored and hence this is the
three-flavor scale).  It is related to the Higgs expectation value
through the one-loop renormalization group equation as (we do not
consider higher loop effects here)
\begin{equation}
  \Lambda_{QCD}^9 = m_c^{2/3} m_b^{2/3} m_t^{2/3} M^5 e^{-8\pi^2/g_s^2(M)}
\end{equation}
where $M$ is some high scale and each quark mass is proportional to
$v$.  The coupling of the Higgs to the proton is given by expanding
the vacuum expectation value as $v \rightarrow v + h$, and hence
\begin{equation}
  y_{pph} = \frac{\partial m_p}{\partial v}
  = \frac{2}{9} \frac{m_p}{v}.
\end{equation}
This allows us to
compute the scattering cross section of the $S$ boson and the
nucleon.  

\begin{figure}[h]
  \centering
  \vspace{.5cm}
  \begin{fmffile}{Sp}
    \begin{fmfgraph*}(100,40) 
      \fmfleft{i1,i2}
      \fmfright{o1,o2}
      \fmf{fermion}{i1,v1,o1}
      \fmflabel{$p,n$}{i1}
      \fmf{dashes}{i2,v2,o2}
      \fmf{dashes,label=$h$}{v1,v2}
      \fmflabel{$S$}{i2}
      \fmfv{label=$-ikv$}{v2}
      \fmfv{label=$-i\frac{2}{9}\frac{m_p}{v}$}{v1}
    \end{fmfgraph*}
   \vspace{.5cm}
 \end{fmffile}
  \caption{Feynman diagrams for the scattering of the $S$-boson off a
    proton or neutron.}\label{fig:Sp}
\end{figure}
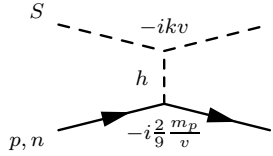

\begin{figure}
\begin{center}
\includegraphics[width=0.7\textwidth]{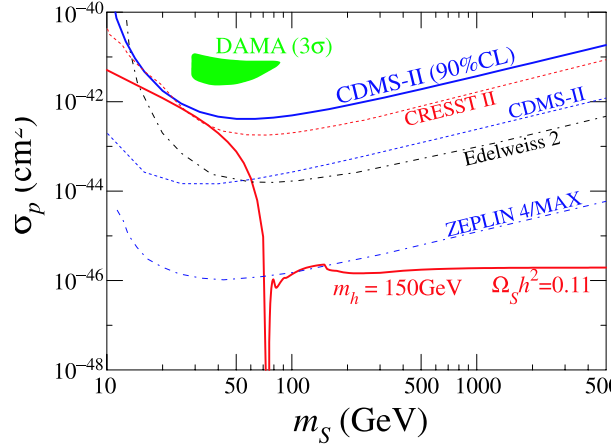}
\end{center}
\caption{The elastic scattering cross section of Dark Matter from nucleons in NMSM, as a function of the Dark Matter particle mass $m_S$ for $m_h=150$~GeV. Note that the region 
$m_S  > 1.8$~TeV is disallowed by the triviality bound on $k$.  Also shown are the experimental bounds from CDMS-II \cite{:2004fq} and DAMA~\cite{Bernabei:2003za}, as well as improved sensitivities expected in the future \cite{dmtools}.  Taken from \cite{Davoudiasl:2004be}. }
 \label{fig:sigmap}
\end{figure}

The result is shown in Fig.~\ref{fig:Sp} as the red solid line.  The
point is that the elastic scattering cross section tends to be {\it
  very}\/ small.  Note that a hypothetical neutrino of the similar
mass would have a cross section of $\sigma_p^\nu \sim G_F^2
m_\nu^2/\pi \sim 10^{-34}$~cm$^2$ which is much much bigger than this.
This is the typical WIMP scattering cross section.  Superimposed is
the limit from the CDMS-II experiment \cite{:2004fq} and hence the
direct detection experiments are just about to reach the required
sensitivity.  In other words, this simple model is completely viable,
and may be tested by future experiments. For the resonance region $m_S
\simeq m_h/2$, the coupling $k$ is very small to keep enough abundance
and hence the direct detection is very difficult.

The future of this field is not only to detect WIMPs but also
understand its identity.  For this purpose, you want to combine the
accelerator data and the direct detection experiments.  The direct
detection experiments can measure the energy deposit and hence the
mass of the WIMP.  It also measures the scattering cross section, even
though it suffers from the astrophysical uncertainty in the estimate
of the local halo density.  On the other hand, assuming $m_S < m_h/2$,
the Higgs boson decays {\it invisibly}\/ $h \rightarrow SS$.  Such an
invisible decay of the light Higgs boson can be looked for at the LHC
using the $W$-fusion process.  Quarks from both sides radiate an
off-shell $W$-boson that ``fuse'' in the middle to produce a Higgs
boson.  Because of the kick by the off-shell $W$-boson, the quarks
acquire $p_T \sim m_W/2$ in the final state and can be tagged as
``forward jets.''  Even though the Higgs boson would not be seen, you
may ``discover'' it by the forward jets and missing $E_T$
\cite{Eboli:2000ze}.  The ILC can measure the mass of the Higgs
precisely even when it decays dominantly invisibly (see, {\it e.g.}\/,
\cite{Weiglein:2004hn}) and possibly its width.  Combining it with the
mass from the direct detection experiments, you can infer the coupling
$k$ and calculate its cosmic abundance.  It would be a very
interesting test if it agrees with the cosmological data $\Omega_M
\approx 0.23$.  If it does, we can claim a victory; we finally
understand what dark matter {\it is}\/!

\subsection{Popular WIMPs}

Superpartners of the photon and $Z$, and neutral Higgs bosons (there
are two of them), mix among each other once $SU(2) \times U(1)$
symmetry breaks.  Out of four such ``neutralino'' states, the lightest
one is often the LSP\footnote{Superpartner of neutrinos is not out of
  the question \cite{Arkani-Hamed:2000bq}. }  and is the most popular
candidate for dark matter in the literature (see, {\it e.g.}\/,
\cite{Ellis:2006jy,Baer:2006id,Trotta:2006ew} for some of the recent
papers).  One serious problem with the supersymmetric dark matter is
that there are many parameters in the model.  Even the Minimal
Supersymmetric Standard Model (MSSM) has 105 more parameters than the
standard model.  It is believed that the fundamental theory determines
all these parameters or at least reduces the number drastically, and
one typically ends up with five or so parameters in the study.
Depending on what parameter set you pick, the phenomenology may be
drastically different.  For the popular parameter set called CMSSM
(Constrained MSSM, also called minimal supergravity or mSUGRA) with
four parameters and one sign, see a recent study in
\cite{Ellis:2003cw}.  I do not go into detailed discussions about any
of them.  I rather mention a few generic points.

First, we {\it do}\/ get viable dark matter candidates from sub-TeV
supersymmetry as desired by the hierarchy problem.  This is an
important point that shouldn't be forgotten.  Second, what exactly is
the mass and composition of the neutralino depends on details of the
parameter set.  The supersymmetric standard model may not be minimal
either; an extension with additional singlet called Next-to-Minimal
Supersymmetric Standard Model (NMSSM) is also quite popular.  Third,
sub-TeV supersymmetry can be studied in great detail at LHC and
(hopefully) ILC, so that we can measure their parameters very
precisely (see \cite{Tsukamoto:1993gt} for the early work).  One can
hope to correlate the accelerator and underground data to fully test
the nature of the dark matter \cite{Baltz:2006fm}.  Fourth, a large
number of particles present in the model may lead to interesting
effects we did not consider in the discussions above.  I'll discuss
one of such effects briefly below.

Universal Extra Dimension (UED) is also a popular model.  I'm sure
Geraldine Servant will discuss dark matter in this model in her
lectures because she is one of the pioneers in this area
\cite{Servant:2002aq}.  Because the LKP is stable, it is a dark matter
candidate.  Typically the first KK excitation of the $U(1)_Y$ gauge
boson is the LKP, and its abundance can be reasonable.  Its prospect
for direct and indirect detection experiments is also interesting.
See a review article that came out after Les Houches
\cite{Hooper:2007qk}.

The most striking effect of having many particle species is the {\it
  coannihilation}\/ \cite{Griest:1990kh}.  An important example in the
case of supersymmetry is bino and stau.  Bino is the superpartner of
the $U(1)_Y$ gaugino (mixture of photino and zino), and its
annihilation cross section tends to be rather small partly because it
dominantly goes through the $P$-wave annihilation.  If, however, the
mass of the stau is not too far above bino, stau is present with the
abundance suppressed only by $e^{-\Delta m/T}$ ($\Delta m =
m_{\tilde{\tau}} - m_{\tilde{B}})$ assuming they are in chemical
equilibrium $\tilde{B} \tau \leftrightarrow \gamma \tilde{\tau}$.
There are models that suggest the mass splitting is indeed small.  The
cross sections $\tilde{B} \tilde{\tau} \rightarrow \gamma \tau$ and
$\tilde{\tau}\tilde{\tau}^* \rightarrow \gamma \gamma, f\bar{f}$ etc
tend to be much larger, the former going through the $S$-wave and the
latter with many final states.  Therefore, despite the Boltzmann
suppression, these additional contributions may win over $\langle
\sigma_{\tilde{B}\tilde{B}} v \rangle$.  There are other cases where
the mass splitting is expected to be small, such as higgsino-like
neutralinos \cite{Mizuta:1992qp}.  In the UED, the LKP is quite close
in its mass to the low-lying KK states and again coannihilation is
important.

\subsection{Indirect Detection Experiments}

On the experimental side, there are other possible ways of detecting
signals of dark matter beyond the underground direct detection
experiments and collider searches.  They are {\it indirect
  detection}\/ experiments, namely that they try to detect
annihilation products of dark matter, not the dark matter itself.  For
annihilation to occur, you need some level of accumulation of dark
matter.  The possible sites are: galactic center, galactic halo, and
center of the Sun.  The annihilation products that can be searched for
include gamma rays from the galactic center or halo, $e^+$ from the
galactic halo, radio from the galactic center, anti-protons from the
galactic halo, and neutrinos from the center of the Sun.  Especially
the neutrino signal complements the direct detection experiments
because the sensitivity of the direct detection experiments goes down
as $1/m_\chi$ because the number density goes down, while the
sensitivity of the neutrino signal remain more or less flat for heavy
WIMPs because the neutrino cross section rises as $E_\nu \propto
m_\chi$.  You can look at a recent review article \cite{Carr:2006cw}
on indirect searches.

\section{Dark Horse Candidates}

\subsection{Gravitino}

Assuming $R$-parity conservation, superparticles decay all the way
down to whatever is the lightest with odd $R$-parity.  We mentioned
neutralino above, but another interesting possibility is that the LSP
is the superpartner of the gravitino, namely gravitino.  Since
gravitino couples only gravitationally to other particles, its
interaction is suppressed by $1/M_{Pl}$.  It practically removes the
hope of direct detection.  On the other hand, it is a possibility we
have to take seriously.  This is especially so in models with
gauge-mediated supersymmetry breaking \cite{Dine:1994vc}.

The abundance of light gravitinos is given by
\begin{equation}
  \Omega_{3/2} h^2 = \frac{m_{3/2}}{2~\mbox{keV}}\, 
\end{equation}
if the gravitinos were thermalized. In most models, however, the
gravitino is heavier and we cannot allow thermal abundance.  The
peculiar thing about a light gravitino is that the longitudinal
(helicity $\pm 1/2$) components have a much stronger interaction
$m_{SUSY}/(m_{3/2} M_{Pl})$ if $m_{3/2} \ll m_{SUSY}$.  Therefore the
production cross section scales as $\sigma \sim m_{SUSY}^2/(m_{3/2}
M_{Pl})^2$, and the abundance scales as $Y_{3/2} \sim \sigma T^3 /
H(T) \sim m_{SUSY}^2 T / (m_{3/2}^2 M_{Pl})$.  Therefore we obtain an
upper limit on the reheating temperature after the inflation
\cite{Moroi:1993mb,deGouvea:1997tn}.  If the reheating temperature is
right at the limit, gravitino may be dark matter.

\begin{figure}[h]
  \centering
  \includegraphics[width=0.5\textwidth]{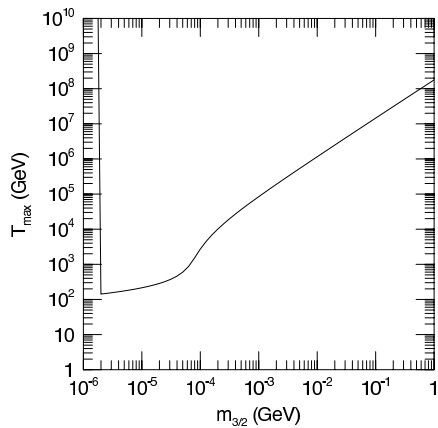}
  \caption{The upper bound on $T_{\rm max}$ as a function of the
    gravitino mass from the requirement that the relic stable
    gravitinos do not overclose the Universe.  Taken from
    \cite{deGouvea:1997tn}.  It assumes $h=1$ and $\Omega_{3/2} h^2 <
    1$ and hence the actual constraint is nearly an order of magnitude
    more stringent than this plot.}
\label{fig:omega_new}
\end{figure}

There is however another mechanism of gravitino production.  The
abundance of the ``LSP''is determined the usual way as a WIMP, while
it eventually decays into the gravitino.  This decay can upset the
Big-Bang Nucleosynthesis \cite{Moroi:1993mb}, but it may actually be
helpful for a region of parameter space to ease some tension among
various light element abundances \cite{Feng:2003uy}.  Note that the
``LSP'' (or more correctly NLSP: Next-Lightest Supersymmetric
Particle) may be even electrically charged or strongly coupled as it
is not dark matter.  When superparticles are produced at colliders,
they decay to the ``LSP'' inside the detector, which escapes and most
likely decays into the gravitino outside the detector.  If the ``LSP''
is charged, it would leave a charged track with anomalously high
$dE/dx$.  It is in principle possible to collect the NLSP and watch
them decay, and one may even be able to confirm the spin $3/2$ nature
of the gravitino \cite{Buchmuller:2004rq} and its gravitational
coupling to matter \cite{Feng:2004gn}.

If the gravitino is heavier than the LSP, its lifetime is calcuated as
\begin{equation}
  \tau(\tilde{G} \rightarrow \gamma \tilde{\gamma})
  = 3.9 \times 10^5 \left( \frac{m_{3/2}}{\rm TeV} \right)^{-3}~\sec.
\end{equation}
It tends to decay after the BBN and upsets its success.  Its
production cross section scales as $\sigma \sim 1/M_{Pl}^2$ and hence
$Y_{3/2} \sim T/M_{Pl}$.  Depending on its mass and decay modes, one
can again obtain upper limits on the reheating temperature.  The case
of hadronic decay for $m_{3/2} \sim 1$~TeV is the most limiting case
that requires $T_{RH} \lesssim 10^{6}$~GeV \cite{Kohri:2005wn},
causing trouble to many baryogenesis models.

\subsection{Axion}

One of the puzzles about the standard model I discussed earlier is why
$\theta \lesssim 10^{-10}$ in QCD.  A very attractive solution to this
problem is to promote $\theta$ to a dynamical field, so that when it
settles to the minimum of the potential, it automatically makes
$\theta$ effectively zero \cite{Peccei:1977hh}.  The dynamical field
is called {\it axion}\/, and couples (after integrating out some heavy
fields) as
\begin{equation}
  {\cal L} = \left( \theta + \frac{a}{f_a} \right)
  \epsilon^{\mu\nu\rho\sigma} {\rm Tr} (G_{\mu\nu} G_{\rho\sigma}).
\end{equation}
Here, $f_a$ is called axion decay constant which has a dimension of
energy.  The potential is given approximately as
\begin{equation}
  V \sim m_\pi^2 f_\pi^2 \left[ 1 - \cos \left( \theta + \frac{a}{f_a}
    \right) \right],
\end{equation}
and indeed the axion field settles to $a = -\theta f_a$ and the
$G\tilde{G}$ term vanishes.  The axion mass is therefore
\begin{equation}
  m_a \approx 6~\mu\mbox{eV} \frac{10^{12}~\mbox{GeV}}{f_a}\ .
\end{equation}
Various astrophysical limits basically require $f_a \gtrsim
10^{10}$~GeV and hence the axion is a very light boson (see, {\it
  e.g.}\/, \cite{Murayama:2000tg}).  Most of these limits come from
the fact that the axion can carry away energy from stars and would
cool them too quickly, such as white dwarfs, red giants, and SN1987A.
Models of such high $f_a$ are called ``invisible axion'' models
because then the axion coupling is very weak to other particles,
avoids these limits, and hence is very difficult to observe.  There
are two popular versions, KSVZ \cite{Kim:1979if,Shifman:1979if} and
DFSZ \cite{Zhitnitsky:1980tq,Dine:1981rt} models.

In the early universe $T \gg $~GeV, the axion potential looks so flat
that it cannot tell where the minimum is.\footnote{This is especially
  true for the axion because its mass originates from the QCD instaton
  effects which are suppressed by powers of the temperature in hot
  thermal bath \cite{Gross:1980br}.}  Therefore we expect it starts
out wherever it finds itself, mostly likely not at the minimum.  The
likely initial misplacement is of the order of $f_a$.  Now we would
like to know what happens afterwards.

Let us consider a scalar field in an expanding universe.  Neglecting
the spatial variation and considering the time dependence alone, the
equation of motion is
\begin{equation}
  \ddot{\phi} + 3 H \dot{\phi} + V'(\phi) = 0.
\end{equation}
For a quadratic potential (mass term) $V(\phi) = \frac{1}{2} m^2
\phi^2$, the equation is particularly simple and homogeneous,
\begin{equation}
  \ddot{\phi} + 3 H \dot{\phi} + m^2 \phi = 0.
\end{equation}
It is useful to solve this equation for a constant (time-independent)
$H$ first.  For Fourier modes $\phi \sim e^{-i\omega t}$ (of course we
take the real part later on), we need to solve
\begin{equation}
  - \omega^2 - 3 i H + m^2 = 0,
\end{equation}
and we find
\begin{eqnarray}
  \omega_\pm &=& \frac{1}{2} \left[ -3iH \pm \sqrt{-9H^2 + 4m^2}
  \right]
  \nonumber \\
  &=& \left\{
    \begin{array}{ccc}
      -3iH, & -i \frac{m^2}{3H} & (H \gg m)\\
      m - i\frac{3}{2} H, & -m - i \frac{3}{2} H & (H \ll m)
    \end{array}
    \right.
\end{eqnarray}
Therefore for $H \gg m$ (early universe), one of the solutions damps
quickly $\phi_+ \sim e^{-3Ht}$, while the other is nearly stationary
$\phi_- \sim e^{-(m^2/3H)t}$.  This is because it is ``stuck'' by the
friction term $-3H \dot{\phi}$.  On the other hand the field
oscillates around the minimum as $\phi = \phi_0 e^{-imt} e^{-3Ht/2}$. 

One can improve this analysis for time-dependent $H(t) = 1/2t$ and
$a(t) \propto t^{1/2}$ in the radiation-dominated univese, by
replacing $e^{-i\omega_\pm t}$ by $e^{-i \int^t \omega_\pm (t') dt'}$
(adiabatic approximation).  When $H \gg m$ ($m t \ll 1$), we find
\begin{eqnarray}
  & &\phi_+ = \phi_0 e^{-3 \int_{t_0}^t H(t') dt'} = \phi_0
  \left( \frac{t_0}{t} \right)^{-3/2}
  = \phi_0 \left( \frac{a_0}{a} \right)^{-3}, \nonumber \\
  & &\phi_- = \phi_0 e^{- m^2 \int_{t_0}^t 2t' dt'/3}
  = \phi_0 e^{-m^2 t^2/3} \approx \phi_0.
\end{eqnarray}
The second one is the solution that is stuck by the friction.  On the
other hand when $H \ll m$,
\begin{eqnarray}
  \phi_\pm = \phi_0 e^{\pm i m t} e^{- \frac{3}{2} \int_{t_0}^t H(t')
    dt'}
  = \phi_0 e^{\pm i mt} \left( \frac{t_0}{t} \right)^{3/4}
  = \phi_0 e^{\pm i mt} \left( \frac{a_0}{a} \right)^{3/2}.
\nonumber  
\end{eqnarray}
The field damps as $a^{-3/2}$, and its energy as $V = m^2 \phi^2/2
\propto a^{-3}$, just like non-relativistic matter.  In fact, a
coherently oscillating homogeneous field can be regarded as a
Bose--Einstein condensate of the boson at zero momentum state.

Therefore, the axion field can sit on the potential and does not roll
down because of the large friction term $-3H\dot{\phi}$ when $H \gg
m$.  On the other hand for later universe $H \ll m$, it oscillates as
a usual harmonic oscillator $e^{\pm i m t}$ and dilutes as
non-relativistic matter.  This is why a very light scalar field can be
a candidate for cold dark matter.  Counterintuitive, but true.  This
way of producing cold dark matter is called ``misalignment
production'' because it is due to the initial misalignment of the
axion field relative to the potential minimum.  Because the amount of
misalignment is not known, we cannot predict the abundance of axion
precisely.  Assuming the misalignment of $O(f_a)$, $f_a \simeq
10^{12}$~GeV is the preferred range for axion dark matter. 

There is a serious search going on for axion dark matter in the halo
of our galaxy.  In addition to the required coupling of the axion to
gluons, most models predict its coupling to photons $a (\vec{E} \cdot
\vec{B})$ of the similar order of magnitude.  The ADMX experiment
places a high-$Q$ cavity in a magnetic field.  When an axion enters
the cavity, this coupling would allow the axion to convert to a
photon, which is captured resonantly by the cavity with a high
sensitivity.  By changing the resonant frequencies in steps, one can
``scan'' a range of axion mass.  Their limit has just reached the KSVZ
axion model \cite{Asztalos:2003px,Duffy:2006aa},\footnote{They ignored
  theoretical uncertainty in the prediction of the axion-to-photon
  coupling, and the KSVZ model is not quite excluded yet
  \cite{Buckley}.} and an upgrade to reach the DFSZ axion model using
SQUID is in the works.  See \cite{Bradley:2003kg} for more on axion
microwave cavity searches.

\subsection{Other Candidates}

I focused on the thermal reclic of WIMPs primarily because there is an
attractive coincidence between the size of annihilation cross section
we need for the correct abundance and the energy scale where we expect
to see new partices from the points of view of electroweak symmetry
breaking and hierarchy problem.  Axion is not connected to any other
known energy scale, yet it is well motivated from the strong CP
problem.  On the other hand, nature may not necessarily tell us a
``motivation'' for a particle she uses.  Indeed, people have talked
about many other possible candidates for dark matter.  You may want to
look up a couple of keywords: sterile neutrinos, axinos, warm dark
matter, mixed dark matter, cold and fuzzy dark matter, $Q$-balls,
WIMPZILLAs, etc.  Overall, the candidates in this list range in their
masses from $10^{-22}$~eV to $10^{22}$~eV, not to mention still
possible MACHOs $\lesssim 10^{-7}M_\odot = 10^{59}$~eV.  Clearly, we
are making progress.

\section{Cosmic Coincidence}

\begin{figure}[h]
  \centering
  \includegraphics[width=0.7\textwidth]{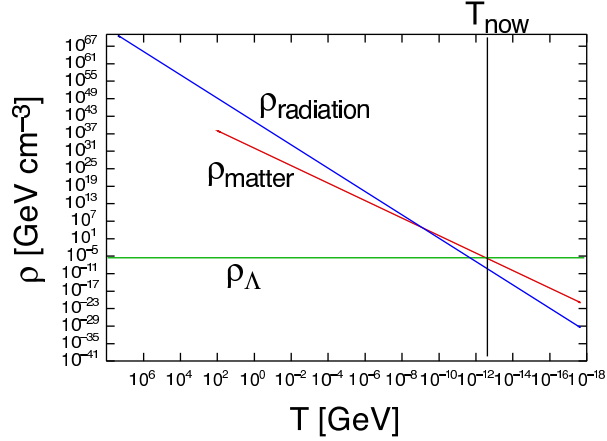}
  \caption{The evolution of radiation, matter, and cosmological
    constant ($\Lambda$) components of the univese as the temperature
    drops over many orders of magnitude.  "Now" is a very special
    moment when matter and $\Lambda$ are almost exactly the same, and
    the radiation is not that different either.}
  \label{fig:coincidence}
\end{figure}

Whenever I think about what the univese is made of, including baryons,
photons, neutrinos, dark matter, and dark energy, what bothers me (and
many other people) is this question: why do they have energy densities
within only a few orders of magnitude?  They could have been many
orders of magnitude different, but they aren't.  This question is
related to the famous ``Why now?'' problem.  The problem is clear in
Fig.~\ref{fig:coincidence}.  As we think about evolution of various
energy densities over many decades of temperatures, why do {\it we}\/
live at this special moment when the dark matter and dark energy
components become almost exactly the same?  I~feel like I'm back to
Ptolemy from Copernicus.  {\it We are special}\/, not in space any
more, but in time.  Is that really so?

In Fig.~\ref{fig:coincidence}, the radiation component goes down as
$T^{-4}$, while the matter $T^{-3}$.  The cosmological constant is by
definition constant $T^0$.  Matter and $\Lambda$ meet {\it now}\/.
When thinking about this problem, it is always tempting but dangerous
to bring ``us'' into the discussion.  Then we will be forced to talk
about conditions for emergence of intelligent lifeforms, which we
don't know very well about.  Instead, it may be better to focus on
physical quantities; namely the triple coincidence problem that three
lines with different slopes seem to more or less meet at a point.  In
fact, dimensional analysis based on TeV-scale WIMP suggests
\begin{equation}
  \rho_{\rm matter} \sim \left(\frac{{\rm TeV}^2}{M_{Pl}}\right)^3 T^3,
\end{equation}
which agrees with $\rho_{\rm radiation} \sim T^4$ at the temperature
$T \sim {\rm TeV}^2/M_{Pl} \approx \mbox{meV} = 10$~K; this is about
now!  In order for the cosmological constant to meet at the same time,
we suspect there is a deep reason
\begin{equation}
  \rho_\Lambda \sim \left(\frac{{\rm TeV}^2}{M_{Pl}}\right)^4.
\end{equation}
Indeed, $\rho_\Lambda^{1/4}$ is observationally about 2~meV, while
${\rm TeV}^2/M_{Pl} \approx 0.5$~meV.  Maybe that is why we see a
coincidence \cite{Arkani-Hamed:2000tc}.  

Actually, an exact coincidence does not leave a window for structure
formation, which requires matter-dominated period.  Fortunately, WIMP
abundance is enhanced by weakness of the annihilation cross section,
which goes like $1/\alpha^2$.  This enhancement of matter relative to
the triple coincidence gives us a window for matter domination and
structure formation.  May be that is why {\it we}\/ seems to be in
this triangle.  But then, why is the baryon component also just a
factor of five smaller than dark matter?   Are they somehow related?

Oh well, we know so little.

\section{Conclusions}

In my lectures, I tried to emphasize that we are approaching an
exciting time to cross new threshold of rich physics at the TeV energy
scale in the next few years at the LHC.  At the same time, the dark
matter of the universe is now established to be not made of particles
we know, requiring physics beyond the standard model.  The main
paradigm for dark matter now is WIMPs, TeV-scale particles produced by
the Big Bang which naturally give the correct order of magnitude for
its abundance.  Even though nature may be tricking us by this
coincidence, many of us (including I) think that there is indeed a new
particle (or many of them) waiting to be discovered at the LHC (or
ILC) that tells us something about the dark side of the universe.  If
this is so, I would feel lucky to be born to this age.

\appendix

\section{Gravitational Lensing}

Gravitational lensing is an important tool in many studies in
cosmology and astrophysics.  In this appendix I introduce the
deflection of light in a spherically symmetric gravitational field
(Schwarzschild metric)

\subsection{Deflection Angle}

Using the Schwarzschild metric ($c=1$)
\begin{equation}
  d s^2 = \frac{r-r_S}{r} \, dt^2 - \frac{r}{r-r_S}\,  dr^2
  - r^2 \, d\theta^2 - r^2 \sin^2\theta \, d\phi^2
\end{equation}
where $r_S = 2 G_N m$ is the Schwarzschild radius.  The
Hamilton--Jacobi equation\footnote{For an introduction to
  Hamilton--Jacobi equations, see
  \url{http://hitoshi.berkeley.edu/221A/classical2.pdf}.} for light in
this metric is 
\begin{eqnarray}
  \lefteqn{
    g^{\mu\nu} \frac{\partial S}{\partial x^\mu}
    \frac{\partial S}{\partial x^\nu} } \label{eq:HJ}\\
  & & 
  \hspace{-.3cm}
  = \frac{r}{r-r_S} \left( \frac{\partial S}{\partial t}\right)^2
  - \frac{r-r_S}{r} \left( \frac{\partial S}{\partial r}\right)^2
  - \frac{1}{r^2} \left( \frac{\partial S}{\partial \theta}\right)^2
  - \frac{1}{r^2 \sin^2\theta} \left( \frac{\partial S}{\partial \phi}\right)^2
  \nonumber \\
  & &
  \hspace{-.3cm}
  =0. \nonumber 
\end{eqnarray}
We separate the variables as
\begin{equation}
  S(t,r,\theta,\phi) = S_1(t) + S_2(r) + S_3(\theta) + S_4(\phi)
\end{equation}
where
\begin{eqnarray}
 \frac{r}{r-r_S} \left( \frac{d S_1}{d t}\right)^2
  - \frac{r-r_S}{r} \left( \frac{d S_2}{d r}\right)^2
  - \frac{1}{r^2} \left( \frac{d S_3}{d \theta}\right)^2
  - \frac{1}{r^2 \sin^2\theta} \left( \frac{d S_4}{d \phi}\right)^2
 \nonumber \\
[.3cm] =0.\hspace{9.5cm}  \nonumber
\end{eqnarray}
Because the equation does not contain $t$ or $\phi$ explicitly, their
functions must be constants,
\begin{eqnarray}
  \frac{d S_1}{d t} &=& -E,\\
  \frac{d S_4}{d\phi} &=& L_z.
\end{eqnarray}
We can solve them immediately as
\begin{eqnarray}
  S_1(t) &=& -Et,\\
  S_4(\phi) &=& L_z \phi.
\end{eqnarray}
Then Eq.~(\ref{eq:HJ}) becomes
\begin{eqnarray}
  \frac{r}{r-r_S} E^2
  - \frac{r-r_S}{r} \left( \frac{d S_2}{d r}(r)\right)^2
  - \frac{1}{r^2} \left( \frac{d S_3}{d \theta}(\theta)\right)^2
  - \frac{1}{r^2 \sin^2\theta} L_z^2
  =0. \nonumber 
\end{eqnarray}
The $\theta$ dependence is only in the last two terms and hence
\begin{equation}
  \left( \frac{d S_3}{d \theta}(\theta)\right)^2
  + \frac{1}{\sin^2\theta} L_z^2 = L^2
\end{equation}
is a constant which can be integrated explicitly if needed.  Without a
loss of generality, we can choose the coordinate system such that the
orbit is on the $x$-$y$ plane, and hence $L_z=0$.  In this case,
$S_4(\phi)=0$ and $S_3(\theta)=L \theta$. Finally, the equation
reduces to
\begin{equation}
  \frac{r}{r-r_S} E^2
  - \frac{r-r_S}{r} \left( \frac{d S_2}{d r}(r)\right)^2
  - \frac{L^2}{r^2}
  =0. \label{eq:HJ3}
\end{equation}
Therefore, 
\begin{equation}
  S_2(r) = \int \sqrt{\frac{r^2}{(r-r_S)^2} E^2 -
    \frac{L^2}{r(r-r_S)}}\ dr. \label{eq:S2}
\end{equation}
Since $S(t,r,\theta,\phi) = S_2(r)-Et+L\theta$, $S_2$ can be regarded
as Legendre transform $S_2(r,E,L)$ of the action, and hence the
inverse Legendre transform gives
\begin{equation}
  \frac{\partial S_2(r,E,L)}{\partial L}=-\theta.
\end{equation}
Using the expression Eq.~(\ref{eq:S2}), we find
\begin{equation}
  \theta(r) = \int_{r_c}^r \frac{Ldr}{\sqrt{E^2 r^4 -  L^2 r(r-r_S)}}\  .
  \label{eq:orbit}
\end{equation}
The closest approach is where the argument of the square root
vanishes,
\begin{equation}
  E^2 r_c^4 - L^2 r_c (r_c-r_S) = 0. \label{eq:closest}
\end{equation}

It is useful to verify that the $m=0$ ($r_S=0$) limit makes sense.
The closest approach is $E^2 r_c^4 - L^2 r_c^2=0$ and hence $r_c =
L/E$, which is the impact parameter.  The orbit Eq.~(\ref{eq:orbit})
is
\begin{equation}
  \theta(r) = \int_{r_c}^r \frac{Ldr}{\sqrt{E^2 r^4-L^2 r^2}}
  = \int_{r_c}^r \frac{r_c dr}{r\sqrt{r^2-r_c^2}}.
\end{equation}
Change the variable to $r=r_c\cosh\eta$, and we find
\begin{equation}
  \theta(r) = \int_{0}^{\eta} 
  \frac{r_c^2\sin\eta d\eta}{r_c\cosh\eta r_c\sinh\eta}
  = \int_0^\eta \frac{d\eta}{\cosh\eta}
  = 2\arctan \tanh \frac{\eta}{2}.
\end{equation}
Hence $\tan\frac{\theta}{2} = \tanh\frac{\eta}{2}$, and
\begin{equation}
  \cos\theta = \frac{1-\tan^2\theta/2}{1+\tan^2\theta/2}
  = \frac{1-\tanh^2\eta/2}{1+\tanh^2\eta/2}
  = \frac{1}{\cosh\eta}
  = \frac{r_c}{r}.
\end{equation}
Therefore $r_c=r\cos\theta$ which is nothing but a straight line.

To find the deflection angle, we only need to calculate the asymptotic
angle $\theta(r=\infty)$.  Going back to Eq.~(\ref{eq:orbit}), we need
to calculate
\begin{equation}
  \theta(\infty) = \int_{r_c}^\infty 
  \frac{Ldr}{\sqrt{E^2 r^4 -  L^2 r(r-r_S)}}\  .
\end{equation}
We would like to expand it up to the linear order in $r_S \ll r_c$.
If you naively expand the integrand in $r_S$, the argument of the
square root in the resulting expression can be negative for
$r=r_c<L/E$.  To avoid this problem, we change the variable to
$r=r_c/x$:
\begin{equation}
  \theta(\infty) = \int_0^1 \frac{Lr_c dx}{\sqrt{E^2 r_c^4-L^2
      r_c(r_c-r_S x)x^2}} \ .
\end{equation}
Using Eq.~(\ref{eq:closest}), we write $E^2 r_c^4$ and obtain
\begin{equation}
  \theta(\infty) = \int_0^1 \frac{r_c dx}{\sqrt{r_c^2(1-x^2) -
      r_cr_S(1-x^3)}} \ .
\end{equation}
Expanding it to the linear order in $r_S/r_c$, we find
\begin{eqnarray}
  \theta(\infty) &=& \int_0^1 \left(\frac{1}{\sqrt{1-x^2}}
    + \frac{(1+x+x^2)r_S}{2(1+x)\sqrt{1-x^2}\ r_c} + O(r_S)^2 \right)
  d x 
 \nonumber \\
 [.2cm]& = & \frac{\pi}{2} + \frac{r_S}{r_c}. 
\end{eqnarray}
The deflection angle is $\Delta\theta = \pi-2\theta(\infty) =
2\frac{r_S}{r_c} = 4G_N m/r_c$.  It is easy to recover $c=1$ by
looking at the dimensions, and we find $\Delta\theta=4G_N m/c^2 r_c$.

It is also useful to know the closest approach $r_c$ to the first
order in $m$.  We expand $r_c$ as $r_c = \frac{L}{E} + \Delta$.  Then
Eq.~(\ref{eq:closest}) gives
\begin{equation}
  4\frac{L^3}{E}\Delta - 2\frac{L^3}{E}\Delta + \frac{E^3}{L} r_S
  + O(r_S)^2 = 0,
\end{equation}
and hence
\begin{equation}
  r_c = \frac{L}{E} -\frac{r_S}{2} + O(r_S)^2.
\end{equation}

\subsection{Amplification in Microlensing}

\begin{figure}[htbp]
  \centering
  \includegraphics[width=0.7\textwidth]{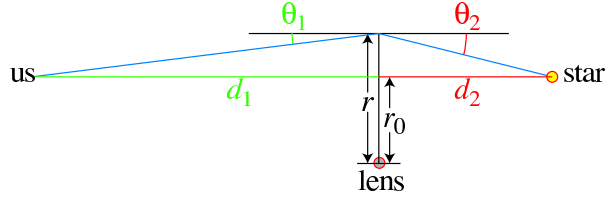}
  \caption{The deflection of light due to a massive body close to the
    line of sight towards a star.}
  \label{fig:microlensing1}
\end{figure}

Once the deflection angle is known, it is easy to work out the
amplification using simple geometric optics.  Throughout the
discussion, we keep only the first order in very small angles.  Just
by looking at the geometry in Fig.~\ref{fig:microlensing1}, the
deflection angle is
\begin{equation}
  \Delta\theta = \theta_1 + \theta_2 =
  \frac{r-r_0}{d_1} + \frac{r-r_0}{d_2} = \frac{4G_N
    m}{r}. \label{eq:Paczynski}
\end{equation}
Here, $r_0$ is the impact parameter.  When $r_0=0$ (exactly along the
line of sight), the solution is simple:
\begin{equation}
  r(r_0=0) = R_0 \equiv \sqrt{4G_N m \frac{d_1 d_2}{d_1+d_2}}.
  \label{eq:R0}
\end{equation}
This is what is called the Einstein radius, $R_0$ in Paczynski's
notation \cite{Paczynski:1985jf}.  For general $r_0$,
Eq.~(\ref{eq:Paczynski}) can be rewritten as
\begin{equation}
  r(r-r_0)-R_0^2 = 0,
\end{equation}
which is Eq.~(1) in the Packzynski's paper.  It has two solutions
\begin{equation}
  r_\pm (r_0) = \frac{1}{2} \left( r_0 \pm \sqrt{r_0^2 + 4 R_0^2}
  \right). \label{eq:r} 
\end{equation}
The solution with the positive sign is what is depicted in
Fig.~\ref{fig:microlensing1}, while the solution with the negative
sign makes the light ray go {\it below}\/ the lens.

\begin{figure}[htbp]
  \centering
  \includegraphics[height=0.4\textheight]{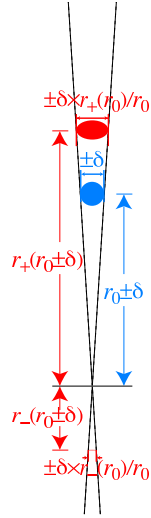}
  \caption{The way the mirror of the telescope appears on the
    deflection plane from the point of view of the star.  For the
    purpose of illustration, we took $R_0=2$, $r_0=3$.}
  \label{fig:microlensing3}
\end{figure}

To figure out the amplification due to the gravitational lensing, we
consider the finite aperture of the telescope ({\it i.e.}\/, the size
of the mirror).  We assume an infinitesimal circular aperture.  From
the point of view of the star, the finite aperture is an image on the
deflection plane of size $\delta$, namely the plane perpendicular to
the straight line from the star to the telescope where the lens is.
The vertical aperture changes the impact parameter $r_0$ to a range
$r_0 \pm \delta$ (size of the mirror is $\delta\times(d_1+d_2)/d_2$).
Correspondingly, the image of the telescope is at $r_\pm(r_0 \pm
\delta) = r_\pm (r_0) \pm \delta \frac{dr_\pm}{dr_0}$.\footnote{Note
  that this Taylor expansion is valid only when $\delta \ll r_0$.  For
  $\delta \sim r_0$, we have to work it out more precisely; see next
  section.}  Using the solution Eq.~(\ref{eq:r}), we find that the
vertical aperture always appears squashed (see
Fig.~\ref{fig:microlensing3}),
\begin{eqnarray}
  \delta \times \left|\frac{dr}{dr_0}\right|
  = \delta \times
  \left| \frac{1}{2} \left( 1 \pm \frac{r_0}{\sqrt{r_0^2+4R_0^2}}\right)
  \right|
  = \delta \times
  \frac{\sqrt{r_0^2+4R_0^2}\pm r_0}{2 \sqrt{r_0^2+4R_0^2}} < \delta .
  \nonumber
\end{eqnarray}
On the other hand, the horizontal aperture is scaled as
\begin{equation}
  \delta \times \frac{r}{r_0}.
\end{equation}
Because the amount of light that goes into the mirror is proportional
to the elliptical aperture from the point of view of the star that
emits light isotropically, the magnification is given by
\begin{eqnarray}
  A_\pm = \frac{r}{r_0} \left| \frac{dr}{dr_0}\right|
  = \frac{(\sqrt{r_0^2+4R_0^2}\pm r_0)^2}{4 r_0 \sqrt{r_0^2+4R_0^2}}
  = \frac{2 r_0^2 + 4R_0^2 \pm 2 r_0\sqrt{r_0^2+4R_0^2}}{4r_0
    \sqrt{r_0^2+4R_0^2}} .
\nonumber    
\end{eqnarray}
The total magnificiation sums two images,
\begin{equation}
  A = A_+ + A_- = \frac{r_0^2 + 2R_0^2}{r_0 \sqrt{r_0^2+4R_0^2}}
  = \frac{u^2+2}{u\sqrt{u^2+4}}
\end{equation}
with $u=r_0/R_0$.\footnote{The singular behavior for $r_0 \rightarrow
  0$ is due to the invalid Taylor expansion in $\delta$.  This is
  practically not a concern because it is highly unlikely that a MACHO
  passes through with $r_0 \lesssim \delta_0$.  Note that the true
  image is actually not quite elliptic but distorted in this case.}
Basically, there is a significant amplification of the brightness of
the star when the lens passes through the line of sight within the
Einstein radius.

\subsection{MACHO search}

We estimate the frequency and duration of gravitational microlensing
due to MACHOs in the galactic halo.  The Large Magellanic Cloud is
about 50kpc away from us, while we are about 8.5kpc away from the
galactic center.  The flat rotation curve for the Milky Way galaxy is
about 220~km/sec (see Fig.~6 in \cite{Raffelt:1997de}).  The Einstein
radius for a MACHO is calculated from Eq.~(\ref{eq:R0}),
\begin{equation}
  R_0 = \sqrt{\frac{4G_N m}{c^2} \frac{d_1 d_2}{d_1+d_2}}
  = 1.24 10^{12}~\mbox{m} \left(\frac{m}{M_\odot}\right)^{1/2}
  \left( \frac{\sqrt{d_1 d_2}}{25\mbox{kpc}}\right).
\end{equation}
To support the rotation speed of $v_\infty = 220$~km/sec in the
isothermal model of halo, we need the velocity dispersion $\sigma =
v_\infty / \sqrt{2}$.  The average velocity transverse to the line of
sight is
\begin{equation}
  \langle v_x^2 + v_y^2 \rangle = 2 \sigma^2 = v_\infty^2.
\end{equation}
The time it takes a MACHO to traverse the Einstein radius is
\begin{equation}
  \frac{R_0}{v_\infty} = 5.6 \times 10^6 \sec
  \left(\frac{m}{M_\odot}\right)^{1/2} \left( \frac{\sqrt{d_1
        d_2}}{25\mbox{kpc}}\right) ,
\end{equation}
about two months for $m=M_\odot$ and $d_1=d_2=25$~kpc.  A microlensing
event of duration shorter than a year can be in principle be
seen.\footnote{MACHO collaboration did even more patient scanning to
  look for microlensing events longer than a year \cite{Allsman:2000kg}.}

The remaining question is the frequency of such microlensing events.
It is the probability of a randomly moving MACHO coming within the
Einstein radius of a star in the LMC.  We will make a crude estimate.
The flat rotation curves requires $\frac{G_N M(r)}{r^2} =
\frac{v_\infty^2}{r}$ and hence the halo density $\rho(r) =
\frac{v_\infty^2}{4\pi G_N r^2}$.  The number density of MACHOs,
assuming they dominate the halo, is then $n(r) =
\frac{v_\infty^2}{4\pi G_N m r^2}$.  Instead of dealing with the
Boltzmann (Gaussian) distribution in velocities, we simplify the
problem by assuming that $\vec{v}^2_\perp = v_x^2 + v_y^2 = \sigma^2$.
From the transverse distance $r_\perp = \sqrt{x^2+y^2}$, only the
fraction $R_0/r_\perp$ heads the right direction for the distance
$\sigma \Delta t$.  Therefore the fraction of MACHOs that pass through
the Einstein radius is
\begin{equation}
  \int_0^{\sigma \Delta t} 2\pi r_\perp dr_\perp \frac{R_0}{r_\perp} 
  = 2\pi R_0 \sigma \Delta t. 
\end{equation}
We then integrate it over the depth with the number density.  The
distance from the solar system to the LMA is not the same as the
distance from the galactic center because of the relative angle
$\alpha = 82^\circ$.  The solar system is away from the galactic
center by $r_\odot = 8.5$~kpc.  Along the line of sight to the LMA
with depth $R$, the distance from the galactic center is given by $r^2
= R^2 + r_\odot^2 - 2R r_\odot \cos\alpha$ with $\alpha=82^\circ$.
Therefore the halo density along the line of sight is
\begin{equation}
  n(r) = 
  \frac{v_\infty^2}{4\pi G_N m (R^2 + r_\odot^2 - 2R r_\odot \cos\alpha)}
\end{equation}
The number of MACHOs passing through the line of sight towards a star
in the LMA within the Einstein radius is
\begin{eqnarray}
  \lefteqn{
    \int_0^{R_{LMC}} dR n(r) 2\pi R_0 \sigma } \nonumber \\
  & & = \int_0^{R_{LMC}} dR \frac{v_\infty^2}{4\pi G_N m (R^2 + r_\odot^2
    - 2R r_\odot \cos\alpha)} 2\pi R_0 \sigma \Delta t
\end{eqnarray}
Pakzynski evaluates the optical depth, but I'd rather estimate a
quantity that is directly relevant to the experiment, namely the rate
of the microlensing events.  Just by taking $\Delta t$ away, 
\begin{eqnarray}
  {\rm rate}
  &=& \int_0^{R_{LMC}} dR \frac{v_\infty^2}{4\pi G_N m (R^2 + r_\odot^2
    - 2R r_\odot \cos\alpha)} 2\pi R_0 \sigma \nonumber \\
  &=& \int_0^{R_{LMC}} dR \frac{v_\infty^2}{R^2 + r_\odot^2
    - 2R r_\odot \cos\alpha}
  \sqrt{\frac{R(R_{LMC}-R)}{G_N mR_{LMC}}}\ \frac{\sigma}{c} \nonumber
  \\
  &=& \frac{v_\infty^2\sigma}{c\sqrt{G_N mR_{LMC}}}
    \int_0^{R_{LMC}}\frac{\sqrt{R(R_{LMC}-R)}\ dR}{R^2 + r_\odot^2
    - 2R r_\odot \cos\alpha}.
\end{eqnarray}
The integral can be evaluated numerically.  For $R_{LMC}=50$~kpc,
$r_\odot = 8.5$~kpc, $\alpha=82^\circ$, Mathematica gives 3.05.  Then
with $\sigma=v_\infty/\sqrt{2}$, $v_\infty = 220$~km/sec, we find
\begin{eqnarray}
  {\rm rate}= 1.69 \times 10^{-13} \sec^{-1}
  \left(\frac{M_\odot}{m}\right)^{1/2} 
  = 5.34 \times 10^{-6} {\rm year}^{-1}
  \left(\frac{M_\odot}{m}\right)^{1/2} .
\nonumber
\end{eqnarray}
Therefore, if we can monitor about a million stars, we may see 5
microlensing events for a solar mass MACHO per year, even more for
ligher ones.

\subsection{Strong Lensing}

\begin{figure}[htbp]
  \centering
  \includegraphics[width=0.5\textwidth]{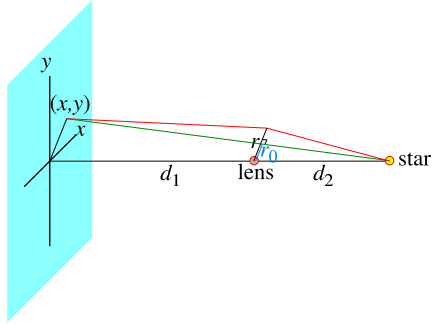}
  \caption{A slightly different coordinate system to work out the
    distortion of images.}
  \label{fig:lensing1}
\end{figure}

Even though it is not a part of this lecture, it is fun to see what
happens when $r_0 \lesssim \delta$.  This can be studied easily with a
slightly tilted coordinates in Fig.~\ref{fig:lensing1}.

\begin{figure}[htbp]
  \centering
  \includegraphics[width=0.3\textwidth]{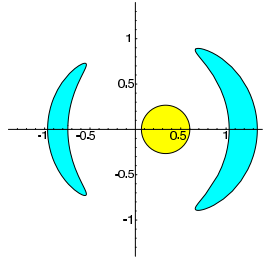}
  \caption{A highly distorted image due to the gravitational lensing.
    Yellow circle is the undistorted image, while the two blue regions
    are the images distorted by the gravitational lensing.}
  \label{fig:lensing2}
\end{figure}

Using this coordinate system, we can draw a circle on the plane
$(x,y)=(x_0, y_0)+\rho(\cos\phi,\sin\phi)$, and the corresponding
image on the deflector plane is $(\tilde{x},\tilde{y})=(\tilde{x}_0,
\tilde{y}_0)+\tilde{\rho}(\cos\phi,\sin\phi) = \frac{d_2}{d_1+d_2}
(x,y)$.  The impact parameter is then $r_0 = \sqrt{\tilde{x}^2 +
  \tilde{y}^2}$ which allows us to calculate $r_\pm (r_0)$ using
Eq.~(\ref{eq:r}) for each $\phi$.  Obviously $\phi$ is the same for
the undistorted and distorted images.  Fig.~\ref{fig:lensing2} shows a
spectacular example with $(x_0,y_0)=(1,0)$,
$\frac{d_2}{d_1+d_2}=\frac{1}{3}$, $\rho=0.8$.  Because $\rho \sim
r_0$, the Taylor expansion does not work, and the image is far from
an ellipse.

\begin{figure}[htbp]
  \centering
  \includegraphics[height=0.4\textwidth]{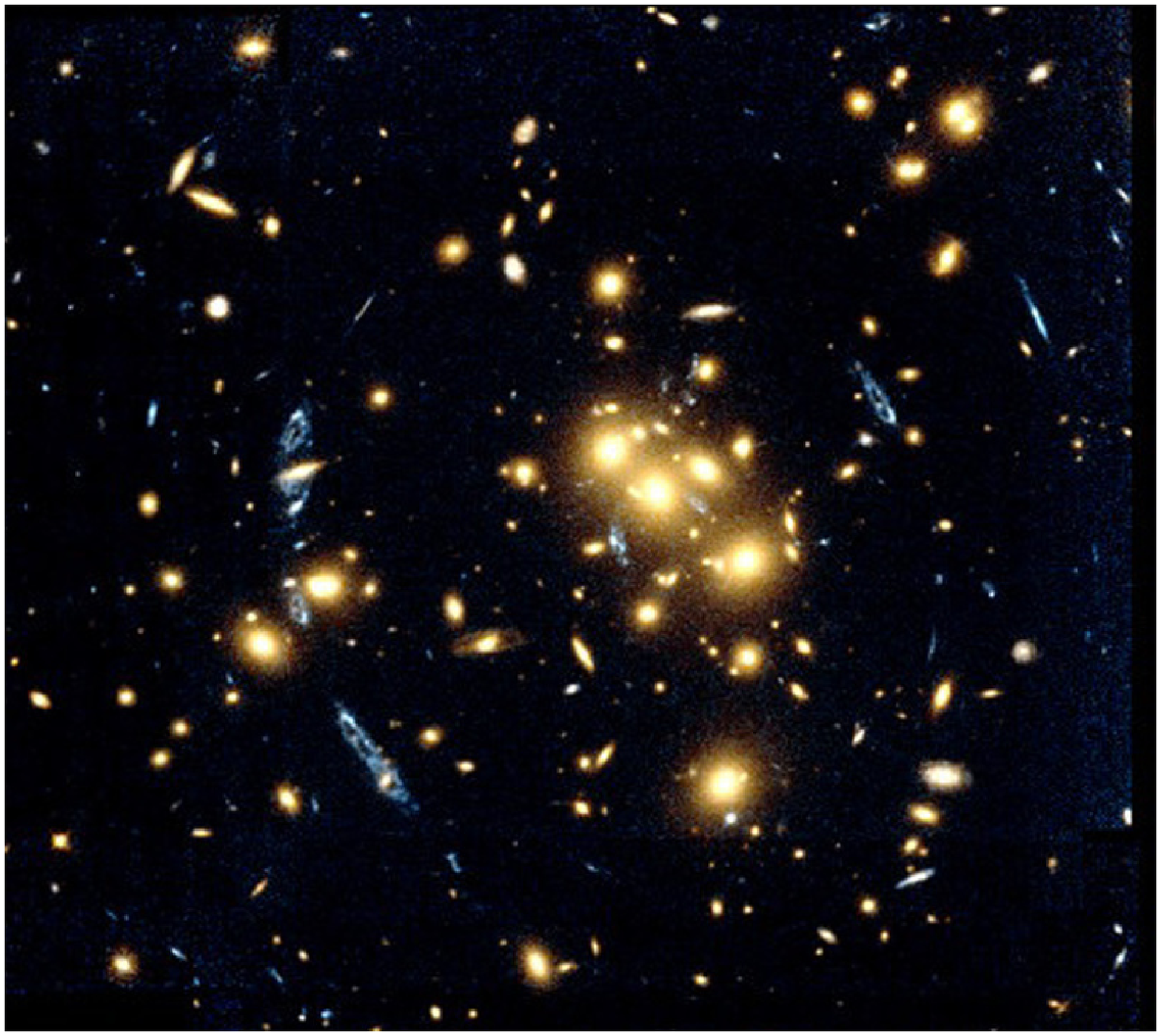}
  \includegraphics[height=0.4\textwidth]{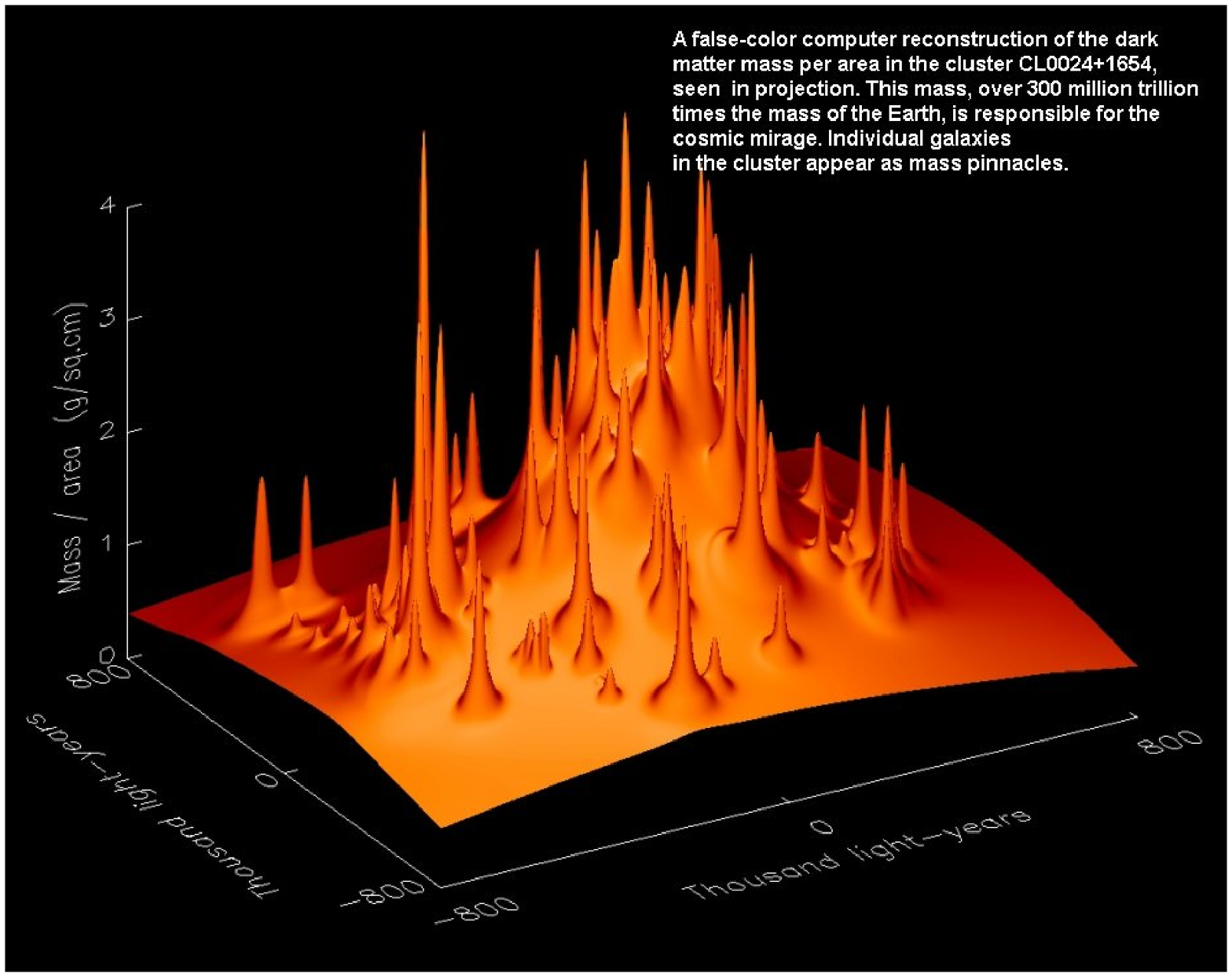}
  \caption{A Hubble Space Telescope image of a gravitational lens
    formed by the warping of images of objects behind a massive
    concentration of dark matter.  Warped images of the same blue
    background galaxy are seen in multiple places.  The detailed
    analysis of lensing effects allows one to map out the mass
    distribution in the cluster that shows a smooth dark matter
    contribution not seen in the optical image.  Taken from
    \cite{Tyson}. }
  \label{fig:hst0024}
\end{figure}

This kind of situation is not expected to occur for something as small
as the mirror of a telescope, but may for something as big as a
galaxy.  When an image of a galaxy is distorted by a concentration of
mass in the foreground, such as a cluster of galaxies, people have
seen spectacular ``strong lensing'' effects, as shown in
Fig.~\ref{fig:hst0024}. 

\section*{Acknowledgements}

I'd like to express my gratitute to the organizers of the school,
Francis Bernardeau and Christophe Grojean.  I am especially indebted
to Christophe for his patience waiting for this contribution.  I also
thank Adam Brown, Edoardo Di Napoli, Alexander Sellerholm, Ethan
Siegel, Daniel Sunhede, Federico Urban, and Gilles Vertongen for our
excursion to T\^ete Rousse.  Matt Buckley kindly read and corrected
the original manuscript.  This work was supported in part by the
U.S. Department of Energy under Contract DE-AC03-76SF00098, and in
part by the National Science Foundation under grant PHY-04-57315.

\end{document}